\documentclass[aps,pra,twocolumn,showpacs,floatfix,superscriptaddress]{revtex4-1}
\usepackage{amsmath}
\usepackage[T1]{fontenc}
\usepackage[dvipdfmx]{graphicx}
\usepackage{bm}
\usepackage{multirow}
\usepackage{braket}
\usepackage{color}
\usepackage{txfonts}
\usepackage{times}
\usepackage{multirow} 
\usepackage{rotating}
\begin{document}

\newcommand{\Yb}[1]{${}^{#1}\text{Yb}$}
\newcommand{\SSS}{${}^{1}\text{S}_{0}$ }
\newcommand{\PPP}{${}^{3}\text{P}_{2}$ }
\newcommand{\PPPS}{${}^{1}\text{P}_{1}$ }
\newcommand{\PPPT}{${}^{3}\text{P}_{1}$ }
\newcommand{\aij}[2]{$\langle \hat{a}^{\dagger}_{#1}\hat{a}_{#2}\rangle$}
\newcommand{\tTOF}{t_{\text{TOF}}}

\allowdisplaybreaks

\title{Experimental Determination of Bose-Hubbard Energies}
\author{Yusuke Nakamura}
\thanks{These two authors contributed equally.}
\affiliation{Department of Physics, Graduate School of Science, Kyoto University, Kyoto 606-8502, Japan}
\author{Yosuke Takasu}
\thanks{These two authors contributed equally.}
\email{Corresponding author}
\affiliation{Department of Physics, Graduate School of Science, Kyoto University, Kyoto 606-8502, Japan}
\author{Jun Kobayashi}
\affiliation{Department of Physics, Graduate School of Science, Kyoto University, Kyoto 606-8502, Japan}
\author{Hiroto Asaka}
\affiliation{Department of Physics, Graduate School of Science, Kyoto University, Kyoto 606-8502, Japan}
\author{Yoshiaki Fukushima}
\affiliation{Department of Physics, Graduate School of Science, Kyoto University, Kyoto 606-8502, Japan}
\author{Kensuke Inaba}
\affiliation{NTT Basic Research Laboratories, NTT Corporation, 3-1, Morinosato Wakamiya, Atsugi, 243-0198, Japan}
\author{Makoto Yamashita}
\affiliation{NTT Basic Research Laboratories, NTT Corporation, 3-1, Morinosato Wakamiya, Atsugi, 243-0198, Japan}
\author{Yoshiro Takahashi}
\affiliation{Department of Physics, Graduate School of Science, Kyoto University, Kyoto 606-8502, Japan}

\date{\today}

\begin{abstract}
We present the first experimental measurement of the ensemble averages of both the kinetic and interaction energies of the three-dimensional Bose--Hubbard model at finite temperature and various optical lattice depths across weakly to strongly interacting regimes, for an almost unit filling factor within single-band tight-binding approximation.
The kinetic energy is obtained through Fourier transformation of a time-of-flight signal, and the interaction energy is measured using a newly developed atom-number-projection spectroscopy technique, by exploiting an ultra-narrow optical transition of two-electron atoms.
The obtained experimental results can be used as benchmarks for state-of-the-art numerical methods of quantum many-body theory.
As an illustrative example, we compare the measured energies with numerical calculations involving the Gutzwiller and cluster-Gutzwiller approximations, assuming realistic trap potentials and particle numbers at nonzero entropy (finite temperature); we obtain good agreement without fitting parameters. 
We also discuss the possible application of this method to temperature estimations for atoms in optical lattices using the thermodynamic relation.
This study offers a unique advantage of cold atom system for ``quantum simulators'', because, to the best of our knowledge, it is the first experimental determination of both the kinetic and interaction energies of quantum many-body system.
\end{abstract}

\maketitle


\section{Introduction}
Ultracold atoms in optical lattices are strongly interacting quantum many-body systems that can be well described by the tight-binding single-band (Bose, Fermi) Hubbard model~\cite{Fisher1989,Jaksch1998}.
The exotic many-body quantum phases of these ``artificial solids" and their phase transition properties have been extensively investigated because of their defect-free lattices and widely tunable experimental parameters, as well as the availability of powerful detection methods~\cite{Bloch2008, BLoch2012}.
An important aim of experiments using artificial solids (so-called ``quantum simulators'') is phase diagram mapping of the fundamental many-body model Hamiltonians.
One of the most interesting problems, which has attracted much attention and has been widely studied, is the quantum phase
transition of ultracold bosonic atoms in a three-dimensional (3D) optical lattice from a superfluid (SF) state to a Mott insulating (MI) state~\cite{Bloch2008}.

The Hamiltonian of the Bose--Hubbard model is given by
\begin{equation}
\hat{\mathcal{H}}=-t\sum_{\langle j, l \rangle}\left(\hat{a}^{\dagger}_j \hat{a}_l +h.c.\right)+\frac{U}{2}\sum_{j} \hat{a}^{\dagger}_j\hat{a}^{\dagger}_j\hat{a}_j\hat{a}_j+\sum_{j}(V_{j}-\mu)\hat{a}^{\dagger}_j\hat{a}_j, ~\label{eq:Hamiltonian}
\end{equation}
where $\hat{a}^{\dagger}_j$, $\hat{a}_j$ are the creation and annihilation operators at site $j$, respectively; $t$ is the tunneling matrix element between nearest-neighbor sites; $U$ is the on-site interaction energy; $\mu$ is the chemical potential; and $V_j$ is the local potential offset at site $j$, which originates from the trap potential and Gaussian envelopes of optical lattice lasers.
Here, $\sum_{\langle j, l\rangle}$ indicates summation over all neighboring sites.
Note that we count only one time per  $\langle j,l \rangle$ pair.

For the Bose--Hubbard system, the competition between the kinetic (atom tunneling) and interaction energies yields a quantum phase transition at low temperature~\cite{Greiner2002}.
In the SF phase, the atoms are spread out over the entire lattice and have long-range phase coherence.
In the MI phase, the atoms are localized at individual lattice sites with integer atom occupancies and have no phase coherence across the entire lattice.
The ratio of $U/t$ determines the quantum phase at zero temperature. 
The system is in the MI or SF phase when $U/t$ > $(U/t)_c$ or $U/t$ < $(U/t)_c$, respectively, with the location of the critical point $(U/t)_c$ depending on the system dimensionality and the filling factor. 
For the 3D homogeneous Bose--Hubbard model at unit filling, $(U/t)_c$ has been numerically calculated to be 29.34(2) using quantum Monte Carlo methods~\cite{Capogrosso-Sansone2007}.

The quantity taken as the experimental observable is important.
Since the first observation of SF-MI transition in 2002~\cite{Greiner2002}, 
the quantities most commonly used to characterize the properties of the quantum states in the Bose-Hubbard system 
have been the visibility and widths of the interference peaks of the time-of-flight (TOF) signals, 
which are sensitive to atomic phase coherence.
These quantities capture the essence of the quantum states. 
In an SF state, the existence of long-range phase coherence over entire lattice sites yields high visibility and narrow widths for the interference peaks in the TOF signal. In contrast, MI state formation is signaled by a decrease in the visibility and broadening of the interference widths, 
resulting from a decrease in the atomic phase coherence.   
Experimental techniques such as noise-correlation measurements~\cite{Folling2005}, quantum gas microscopy~\cite{Bakr2010}, and radio-frequency (RF)~\cite{Campbell2006} and laser spectroscopy~\cite{Kato2016} are used to probe the phase coherence, density-density correlation, and atom number distribution, respectively.
%
%

The most important quantity governing the quantum phase at thermal equilibrium is the Hamiltonian. 
However, despite its crucial importance, there are no reports of systematic measurement of the energy terms in the Hamiltonian; i.e., the ensemble averages of both the kinetic and interaction terms, the competition of which induces the SF-MI quantum phase transition.
The lack of such experiments is partly because no established experimental methods or protocols are known to accurately evaluate the ensemble averages of the kinetic and interaction terms.

Here, we present, to our best knowledge, the first comprehensive measurements of the ensemble averages of both the kinetic and interaction terms at finite temperature and various optical lattice depths, for 3D Bose--Hubbard model with an almost unit filling factor within single-band tight-binding approximation.
We establish a protocol to accurately extract the ensemble average of the kinetic term from the TOF signal, with careful consideration of the finite TOF effect and inter-atomic interaction effect.
We also develop a new method of atom-number-projection spectroscopy, 
which enables direct measurement of the number distributions of multiply occupied sites at any optical lattice depth 
and, hence, accurate evaluation of the ensemble average of interaction terms 
across the weakly to strongly interacting regimes.
Excellent resolution that allows different site-occupancies to be distinguished is obtained by exploiting an ultra-narrow optical transition between the electronic states of $^1\text{S}_0$ and $^3\text{P}_2$, which have quite different on-site interactions in the case of the two-electron atoms of ytterbium (Yb) (see also Appendix~\ref{ap:energy}).
Different from  the standard quantum gas microscopy method, which detects the parity of the atom number at a site due to the pairwise loss of atoms induced by light-assisted collision during fluorescence imaging~\cite{Bakr2009}, 
our atom-number-projection spectroscopy technique can detect {\it any} atom number at an {\it n}-occupied site.
We experimentally examine occupancy-dependent properties such as the finite lifetime and transition probability in order to accurately evaluate the total atom number at the $n$-occupied sites. 
We experimentally determine the kinetic and interaction terms $\langle \hat{K}\rangle=\sum_{\langle j, l \rangle}(\langle \hat{a}^{\dagger}_j \hat{a}_l \rangle+c.c.)$
and $\langle \hat{G}\rangle=\sum_{j} \langle \hat{a}^{\dagger}_j\hat{a}^{\dagger}_j\hat{a}_j\hat{a}_j \rangle$, respectively,
and use the numerical values of the $t(V_0)$ and $U(V_0)$ parameters reported in Ref.~\cite{Krutitsky2016} ($V_0$ is the optical lattice depth).

Using these methods, the ensemble averages of the kinetic and interaction terms are successfully obtained at finite temperature and various optical lattice depths.
These results can be used as benchmarks in state-of-the-art numerical methods pertaining to quantum many-body theory.
In this work, we compare the measured energies with numerical calculations involving Gutzwiller and cluster-Gutzwiller methods at nonzero entropy (finite temperature).
The trap potentials and particle numbers used in the calculations are identical to those of the experiments, and we obtain good agreement without fitting parameters. 
We also discuss application of this experimental method to temperature estimations for atoms in optical lattices using the thermodynamic relation.



This paper is organized as follows: In Sec.~\ref{sec:setup}, we explain our experiment setup and procedure.
The method for measuring the kinetic (interaction) energy is presented in Sec.~\ref{sec:kinetic} (Sec.~\ref{sec:interaction}).
We discussed a possibility of measuring ensemble average of potential energy term in Sec.~\ref{sec:potential}.
In Sec.\ref{sec:energy}, we present our main experimental results, including the kinetic an the interaction energies.
We compare the measured energies with numerical calculations involving Gutzwiller and cluster-Gutzwiller methods at finite temperature in Sec.~\ref{sec:theory}.
Section~\ref{sec:conclusion} is devoted to conclusions and further prospects.

\section{Basic Experiment Setup and Procedure} \label{sec:setup}

We briefly describe the basic experiment setup and procedure here. Further details are given in Appendix~\ref{ap:setup}.
\subsection{Atom preparation}
\begin{figure}[tb]
	\includegraphics[width=9cm]{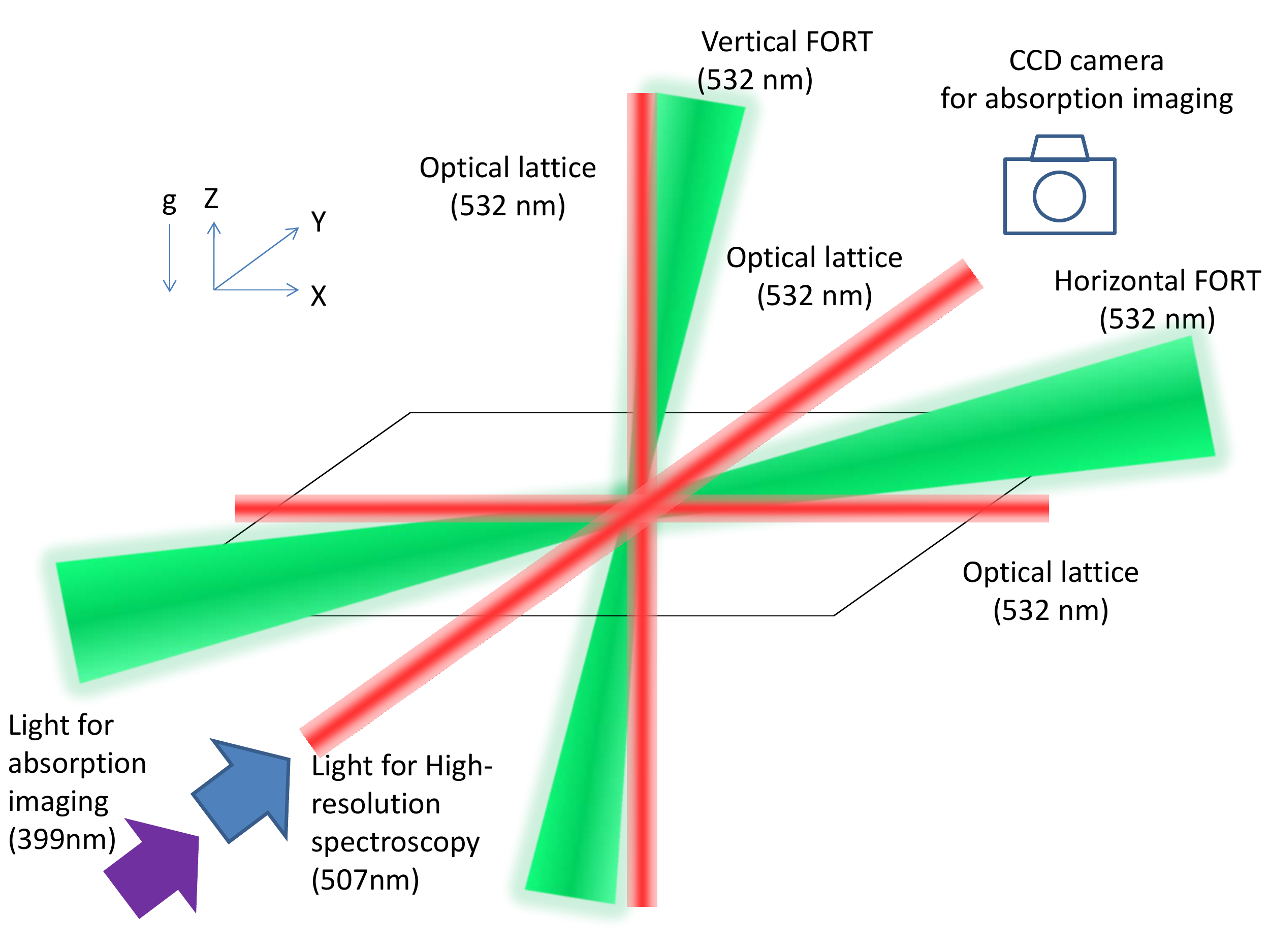}
	\caption{(color online) Schematic view of setup. Only the FORT beams and optical lattice beams, the probe light for absorption imaging, and the excitation light for high-resolution spectroscopy are shown.\label{fig:setup}}
\end{figure}

Our experiment began with magneto-optical trapping of $^{174}$Yb atoms from an atomic oven.
Evaporative cooling was performed using a crossed-beam  optical far-off resonant trap (FORT) geometry formed by two orthogonal horizontal and vertical FORT laser beams of 532-nm wavelength with elliptical laser-beam waists (see Fig.~\ref{fig:setup}).

After preparation of the \Yb{174} Bose-Einstein condensate (BEC), we adiabatically ramped up a 3D cubic optical lattice generated by three orthogonal, retro-reflected laser beams also having 532-nm wavelength and propagating along the X-, Y-, and Z-axes.
The number of atoms before loading onto the optical lattice was stabilized from $1.4$ $\times$ $10^4$ to $1.8$ $\times$ $10^4$.

\begin{figure}[tb]
	\includegraphics[width=9cm]{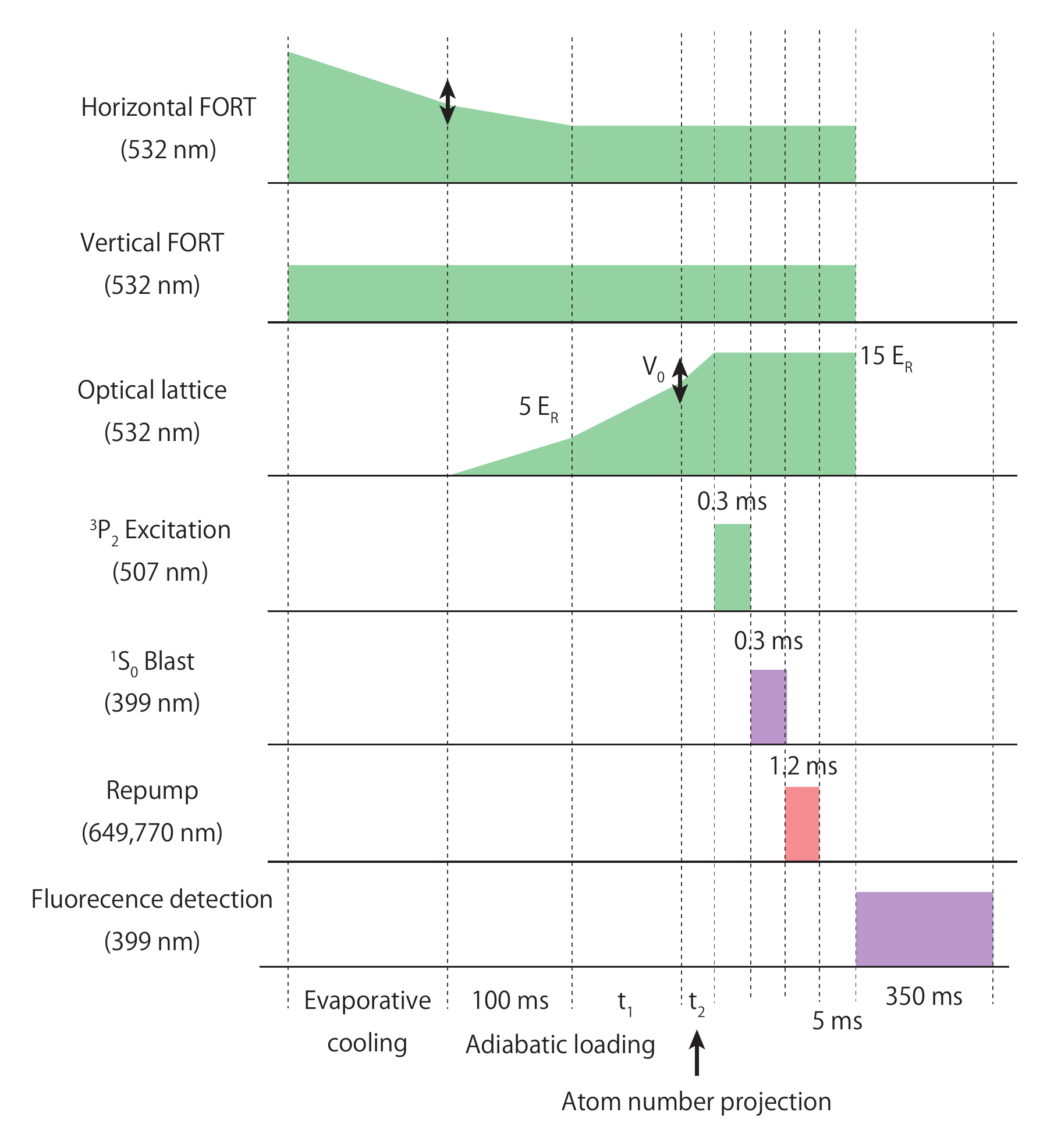}
	\caption{(color online) Schematic time sequence (not scaled) for high-resolution spectroscopy, where $t_1=10(V_0/E_R-5)$ [ms] and $t_2=0.01|V_0/E_R-15|$ [ms]. The double-sided arrows indicate variable parameters (see text for details).\label{fig:sequence-fl}}
\end{figure}

\begin{figure}[tb]
	\includegraphics[width=9cm]{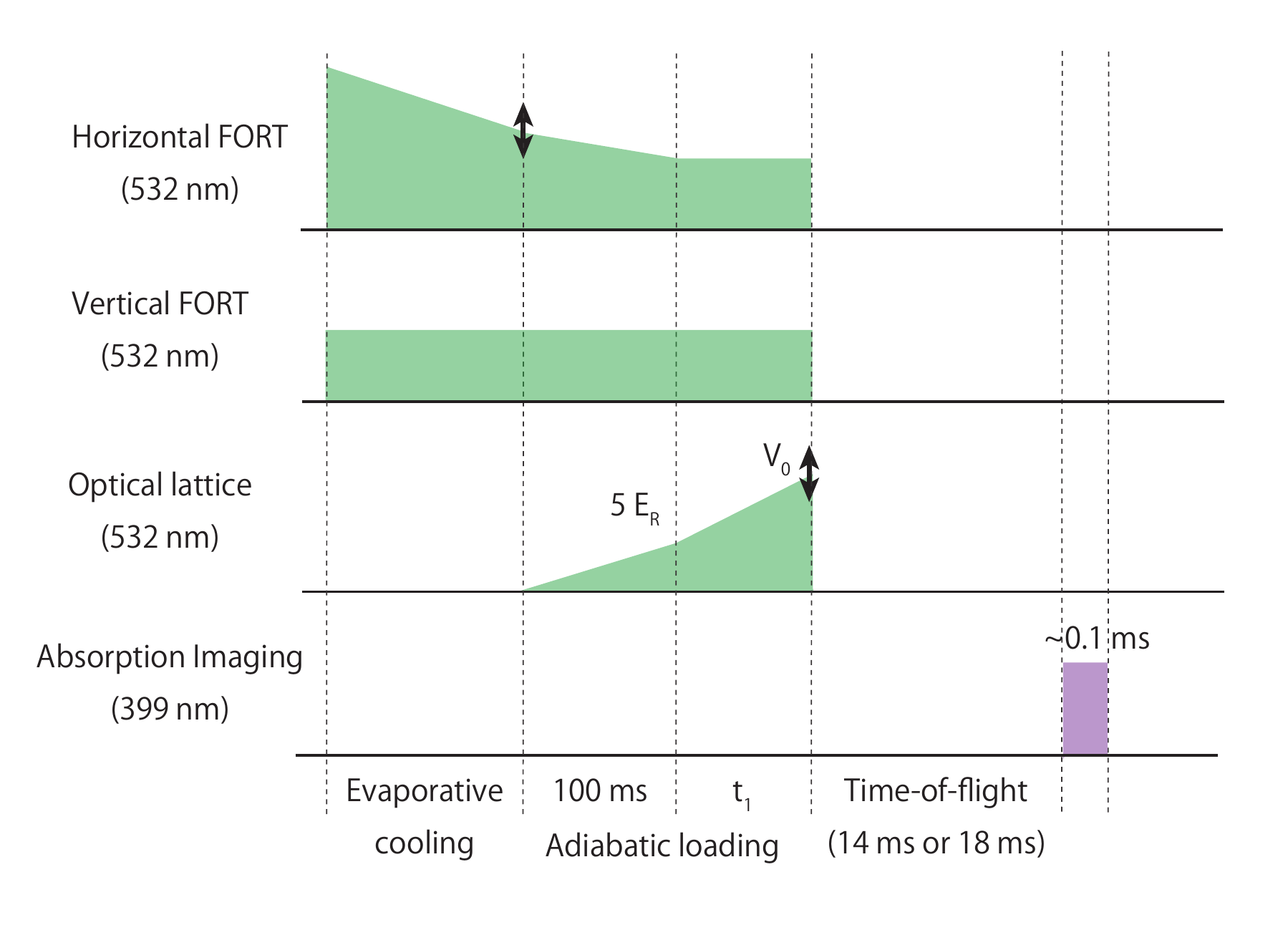}
	\caption{(color online) Schematic time sequence (not scaled) for TOF measurement, where $t_1=10(V_0/E_R-5)$ [ms] and $t_2=0.01|V_0/E_R-15|$ [ms]. The double-sided arrows indicate variable parameters (see text for details).\label{fig:sequence-ab}}
\end{figure}

The experimental procedures for the high-resolution spectroscopy and TOF measurements, including atom loading onto the optical lattice, are shown in Figs.~\ref{fig:sequence-fl} and ~\ref{fig:sequence-ab}, respectively.
In the first 100 ms of loading, the optical lattice depth was increased to 5 $E_R$, where the recoil energy $E_R=h^2/(2m\lambda_L^2)$, with $h$ being the Planck constant and $\lambda_L$ the optical lattice wavelength (532 nm).
Then, we increased the final lattice depth of $V_0$ in $10(V_0/E_R-5)$ ms, with the FORT powers being kept constant.

\subsection{Preparation of various atomic entropies}
One of the important considerations in our experiment was preparation of cold atoms with various atomic entropies in the same experiment setup.
We controlled the atomic entropy by changing the FORT depth in the final stage of evaporative cooling.
Because the FORT depth depends on the horizontal FORT power, we in fact controlled the final horizontal FORT power in this manner.
However, the trap frequencies also depend on the FORT power; therefore, we changed the horizontal FORT power during adiabatic loading onto the optical lattice in the first 100 ms (see Figs.~\ref{fig:sequence-fl} and ~\ref{fig:sequence-ab}).

Because direct measurement of the atomic entropy in the optical lattice is difficult,
we estimated this property from the initial atomic entropy and heating during lattice loading.
The initial entropy $S_1$ in a FORT harmonic trap is~\cite{Pethick_Smith}
\begin{equation}
S_1=4N_1k_B\frac{\zeta(4)}{\zeta(3)}\left(\frac{T}{T_c}\right)^{3},
\end{equation}
where $N_1$ is the atom number; $T$ is the atomic temperature in the FORT, which can be directly measured via a TOF method; $T_c$ is the critical temperature; $\zeta(z)$ is the zeta function; and $k_B$ is the Boltzmann constant. Here,
\begin{equation}
k_B T_c= \hbar \bar{\omega} \left(\frac{N}{\zeta(3)}\right)^{1/3},
\end{equation}
where $\bar{\omega}$ is the geometric mean of the three trap frequencies and $\hbar$ is the Planck constant divided by $2\pi$.

To estimate the additional atomic heating during loading onto the optical lattice, we measured the entropy $S_2$ and atom number $N_2$ after adiabatically ramping down the optical lattice in reverse order (see Appendix~\ref{ap:setup}).
We assumed that the entropy per atom in the optical lattice, $s_{\text{OL}}$, was written as in Eq.~(\ref{eq:entropy_OL}), 
using the entropy before (after) loading onto the optical lattice $S_{\text{1}}$ ($S_{\text{2}}$):
\begin{equation}
s_{\text{OL}}=\frac{1}{2}\left(\frac{S_1}{N_1}+\frac{S_2}{N_2}\right). \label{eq:entropy_OL}
\end{equation}
We obtained the atomic entropy $s_{\text{OL}}$ 
by taking five TOF images and calculating each atomic entropy; these values were then averaged. 

\section{Method for Measuring Kinetic-Term Ensemble Average: Fourier Transformation of TOF Signal} \label{sec:kinetic}

Here, we present a method for obtaining the ensemble average of the first term in Eq. (\ref{eq:Hamiltonian}) (the kinetic term) $-t\langle \hat{K}\rangle$.
We found that  $\langle \hat{K}\rangle$ can be simply measured from  TOF images.
The atomic-density distribution $n_{\text{TOF}} (\mathbf{r}_{\text{TOF}})$ after the TOF $t_{\text{TOF}}$ is given by~\cite{Pedri2001, Gerbier2008}
\begin{equation}
	n_{\text{TOF}}(\mathbf{r}_{\text{TOF}})=\left(\frac{m}{\hbar t_{\text{TOF}}}\right)^3 \left|\tilde{w}_0(\mathbf{k}_{\text{TOF}})\right|^2 S(\mathbf{k}_{\text{TOF}}),
\end{equation}
where $m$ is the atom mass, $\tilde{w}_0 (\mathbf{k}_{\text{TOF}})$ is the Fourier transformation of the Wannier function in the lowest Bloch band $w_0(\mathbf{r})$, and $\mathbf{k}_{\text{TOF}}$ = $m\mathbf{r}_{\text{TOF}}/\hbar t_{\text{TOF}}$.
The structure factor $S(\mathbf{k}_{\text{TOF}})$ is expressed as
\begin{equation}
	S(\mathbf{k}_{\text{TOF}})=\sum_{j,l}e^{i \mathbf{k}_{\text{TOF}}\cdot \left(\mathbf{r}_j-\mathbf{r}_l\right)-i\left(\frac{m}{2\hbar t_{\text{TOF}}}\right)(\mathbf{r}^2_j-\mathbf{r}^2_l)} \langle \hat{a}^{\dagger}_j \hat{a}_l \rangle, \label{eq:TOF}
\end{equation}
where $\mathbf{r}_j$ indicates the site position with index $j$ in the optical lattice and $\langle \cdot\rangle$ represents the ensemble average.
The second term in the exponential, $\exp[-im(\mathbf{r}_j^2-\mathbf{r}_l^2)/(2\hbar t_{\text{TOF}})]$, 
introduces the effect of the finite TOF.  
This term corresponds to the quadratic term in the Fresnel approximation of near-field optics~\cite{Toth2008}.

Details of our derivation are given in Appendix~\ref{ap:TOF}.
Here, for simplicity, we first consider the one-dimensional case and ignore the finite TOF effect.
Equation~(\ref{eq:TOF}) is then expressed as
\begin{equation}
	S(k_x)=\sum_{j,l}e^{i k_x\left(x_j-x_l\right)} \langle \hat{a}^{\dagger}_j \hat{a}_l \rangle. \label{eq:TOF1}
\end{equation}
Note that we omit the ``TOF'' label for simplicity in this section.
We assume that \aij{j}{l} = \aij{l}{j} and
\begin{equation}
S(k_x)=\sum_{j,l}\langle \hat{a}^{\dagger}_j \hat{a}_l \rangle\cos\left[k_x\cdot \left(x_j-x_l\right)\right].
\end{equation}
Next, we define the kinetic energy 
$-t\langle\hat{K}\rangle_x=-t\sum_{\left<j,l\right>}(\langle\hat{a}_j^{\dagger}\hat{a}_l\rangle+c.c.)=\sum E(k_x)\langle\hat{c}^{\dagger}(k_x)\hat{c}(k_x)\rangle$,
where
$\hat{c}(\textbf{k})$ and $\hat{c}^{\dagger}(\textbf{k})$ are the annihilation and creation operators of the Bloch states and
$\hat{c}(\textbf{k})=1/\sqrt{N_L}\sum_{j}\hat{a}_j \exp(i\textbf{k}\cdot\textbf{r}_j)$.
The quasi-momentum $k_x$ runs over the first Brillouin zone only and satisfies the periodic boundary condition
$
k_x=2\pi n_x/(N_{Lx}d_{\text{lat}}), (n_x=0,\pm1\pm2,\cdots).
$
Here, $N_{Lx}$ is the number of lattice sites along the X-axis and 
$d_{\text{lat}}$ is the lattice spacing (266 nm). 
We straightforwardly obtain $\sum\langle \hat{a}^{\dagger}_{j}\hat{a}_{j+1}\rangle$ through a Fourier transformation of $S(k_x)$ in the first Brillouin zone, such that
\begin{eqnarray}
\langle\hat{K}\rangle_x&=&2\sum_j\langle \hat{a}^{\dagger}_{j}\hat{a}_{j+1}\rangle=2\sum_j\langle \hat{a}^{\dagger}_{j}\hat{a}_{j-1}\rangle \nonumber \\ 
&=&\frac{d_{\text{lat}}}{\pi}\int_{-\pi/d_{\text{lat}}}^{\pi/d_{\text{lat}}}  d k_x S(k_x) \cos \left(d_{\text{lat}}k_x\right), \label{eq:FT}
\end{eqnarray}
where
$S(k_x)=N_{Lx}\langle\hat{c}^{\dagger}(k_x)\hat{c}(k_x)\rangle$.
Equation~(\ref{eq:FT}) implies that the energy of the lowest Bloch band is $E(k_x)=-2t\cos(d_{\text{lat}}k_x)$.
To our best knowledge, the above Eq.~(\ref{eq:FT}) has not been explicitly reported to date, despite its importance and simplicity.
In the present work, this simple relation allows us to successfully evaluate the kinetic energy from experimental observation. 

\begin{figure}[tb]
	\includegraphics[width=9cm]{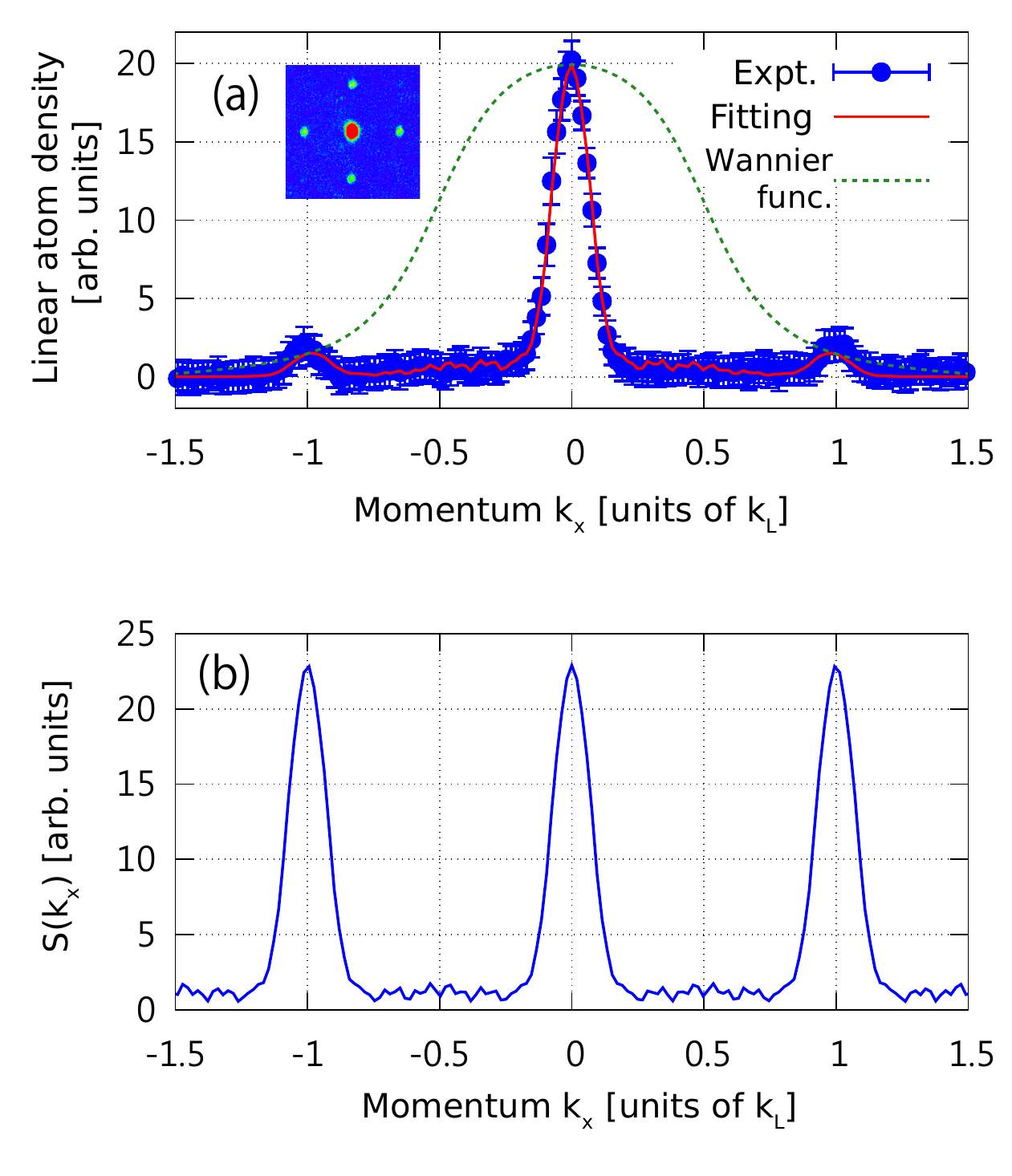}
	\caption{(color online) (a) Linear atom density of TOF signal integrated along vertical axis. Inset: TOF image with identical parameters. The TOF was 14 ms and ten images were averaged. The lattice depth was 5$E_R$ and the atomic entropy was 0.04 $k_b$. The fitting results using the $S(k_x)|\tilde{w}(k_x)|^2$ function (red solid line) and squared Wannier function $|\tilde{w}(k_x)|^2$ (dotted green line) are also shown. Here, $k_L=\pi/d_{\text{lat}}$. (b) $S(k)$ obtained by fitting data shown in (a).\label{fig:tof}}
\end{figure}

In the experiment, we obtained a two-dimensional (2D) atomic-density distribution $I(x,z)$, because the TOF signal was integrated in the probe direction (which we took to be the Y-axis).
From the atomic linear densities along the X- and Z-axes, 
we obtained $S(k_x)$ and $S(k_z)$ by fitting of the $S(k) |\tilde{w}(k)|^2$ function, 
where the Wannier function $\tilde{w}(k)$ was obtained by numerically calculating the lowest band of the optical lattice for non-interacting atoms.
We consider the structure factor of the form $S(k)=\sum_{\alpha=0}^{19}A_\alpha\cos(\alpha kd_{\text{lat}})$, which is depicted in Fig.~\ref{fig:tof} 
($A_\alpha$ are fitting parameters).
Then, we obtained the ensemble averages of the kinetic energy $-t\langle\hat{K}\rangle_x$ ($-t\langle\hat{K}\rangle_z$) from $S(k_x)$ ($S(k_z)$), 
and assumed that the total-ensemble average of the kinetic term $-t\langle \hat{K}\rangle$ was
\begin{equation}
-t\langle\hat{K}\rangle=-\frac{3}{2}t\left(\langle\hat{K}\rangle_x+\langle\hat{K}\rangle_z\right).
\end{equation}

Note that it is possible to obtain non-local atomic correlations $\sum_{j} \langle \hat{a}^\dagger_j \hat{a}_{j+n}\rangle$ ($n$ = $2,3,\cdots$) using the method shown here.
As a demonstration, we directly determine the coherence length in Appendix~\ref{ap:other}.
In addition, the 2D atomic correlation can be obtained using the 2D Fourier transformation.


\subsection{Effect of finite TOF}

The effect of a finite TOF arises from the $\exp[-im(\mathbf{r}_j^2-\mathbf{r}_l^2)(2\hbar t_{\text{TOF}})]$ term in Eq.~(\ref{eq:TOF}).
Instead of adding this effect to TOF image in order to directly compare the experimental results~\cite{Vidmar2015}, we experimentally evaluated the total site number along the $\alpha$ axis $N_{L\alpha}$ ($\alpha=x, z$) in order to remove this effect.
Details are given in Appendix~\ref{ap:TOF}.
The basic concept is that it is possible to evaluate the true value (i.e., the infinite TOF) from experimental measurements with several TOFs through extrapolation.
In this work, we measured the atom correlation of $\langle \hat{K}\rangle_{\alpha}(t_{\text{TOF}}, \Delta l)=\sum_j\langle \hat{a}^{\dagger}_{j}\hat{a}_{j+\Delta l}\rangle$, where $\Delta l$ was 1, 2, 3, 4 and $t_{\text{TOF}}$ was set to TOFs of 14 and 18 ms.
Then, $\langle \hat{K}\rangle_{\alpha}(\infty, \Delta l)$ ($\Delta l$ = 1, 2, 3, 4) and the total site number along the $\alpha$ axis $N_{L\alpha}$ were obtained through fitting using equations similar to Eq.~(\ref{eq:correction}).
For a TOF of 14 (18) ms, a reduction of 6 (4) \% in the value of $\sum_j\langle \hat{a}^{\dagger}_{j}\hat{a}_{j+1}\rangle$ from its value for an infinite TOF was estimated.
Note that this finite TOF effect was experimentally checked using datasets for various TOFs, with the other experimental parameters unchanged, and the validity of our result was confirmed (see Appendix~\ref{ap:TOF}).
Because the correction of the finite-TOF effect is model-dependent, so we consider estimated deference between the values before and after the correction as a systematic error.

It is noted that the finite-TOF effect scales as $\exp\left(-imd_{\text{lat}}^2N_L^2\right)$ for {\it whole} sites.
However, it scales as $\exp\left(-imd_{\text{lat}}^2N_L\right)$ for neighbor sites, which contribute to the kinetic energy, and resulted in highly-suppressed finite-TOF effect for measurement of atom correlation in the neighbor sites.
The corrections itself are estimated to be within 6\% and therefore the selection of the assumption of the density function is not so critical.

\subsection{Effect of inter-atomic interaction during TOF}

The discussion above is based on the Wannier states of non-interacting atoms.
Here we discuss the effect of inter-atomic interaction during TOF.
First is the validity of the Wannier function numerically calculated.
The kinetic energy of non-interacting atoms in the lowest band of the optical lattice is estimated to be $\hbar \omega_L$, where $\omega_\text{L}$ is the oscillation frequency at the bottom of the lattice potential~\cite{Gerbier2008}.
The ratio of the interaction energy  $Un(n-1)/2$ to the kinetic energy  $\hbar \omega_L$, i.e.,
\begin{equation}
\eta=\frac{Un(n-1)}{2\hbar \omega_\text{L}},
\end{equation}
determines the relative importance of the inter-atomic interaction during the TOF. 
The ratio $\eta$ was mostly far lower than 1 under our experimental conditions.
Our calculations indicate that $\eta$ takes the maximum value of 0.25 at 7$E_R$ depth in the case of triple occupancy, $n=3$, for which the population fraction is less than 0.1 (see Sec.~\ref{sec:energy}).
Therefore, the effect of the inter-atomic interaction was negligible in our experiment.

The second possible influence of the inter-atomic interaction on the TOF measurements is the conversion from the interaction energy, to the kinetic energy after release from the optical lattice~\cite{Kashurnikov2002}.
We discuss it in Sec.\ref{sec:discussion}.

\section{Method to Measure Ensemble Average of Interaction Term: Atom-Number-Projection Spectroscopy} \label{sec:interaction}
\begin{figure}[tb]
	\includegraphics[width=9cm]{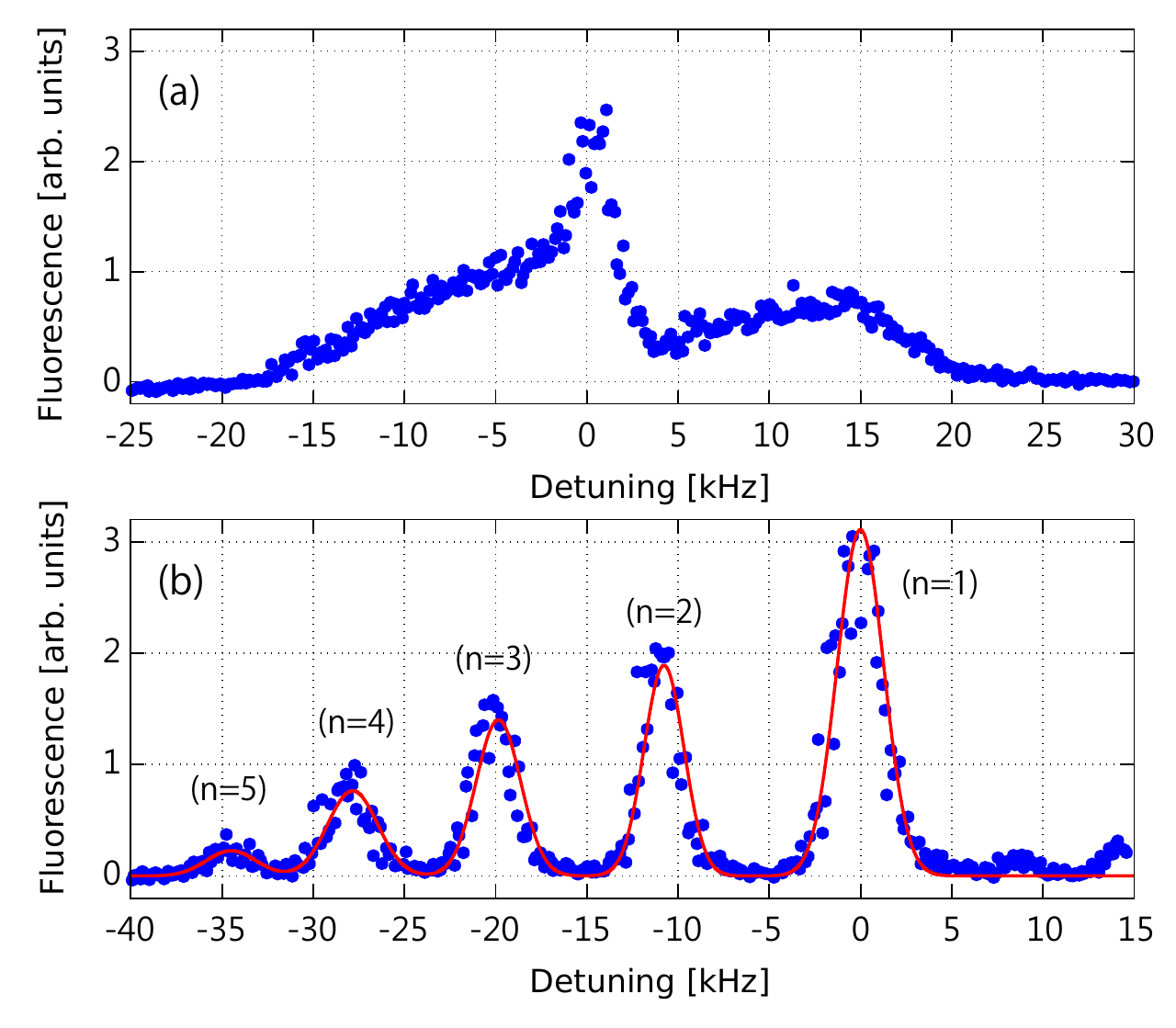}
	\caption{(color online) Atom-number-projection spectroscopy. 
Three scans are superimposed after the long-term laser frequency drift is compensated. 
(a) Single, broad spectra of coexisting SF and normal components are observed at a small optical lattice depth (5$E_R$) without the projection method. (b) Site-occupancy-resolved spectra can be obtained in the small optical lattice (5$E_R$) using the projection method of a sudden increase to 15$E_R$ in 0.1 ms. The solid red line is a fitting as a guide for the eye. Up to five-body occupied sites are observed. \label{fig:projection}}
\end{figure}

We have developed a new method of atom-number-projection spectroscopy, 
which enables direct measurement of the number of multiply occupied sites at {\it any} optical lattice depth and, hence, accurate evaluation of the interaction term {\it across the weakly to strongly} interacting regimes.
%
One may wonder whether such a new method is truly necessary, because information on the numbers of {\it n}-occupied sites is straightforwardly obtained through high-resolution spectroscopy for the atoms in a {\it deep} optical lattice. 
In fact, site-occupancy-resolved spectra in a deep optical lattice have already been reported in~\cite{Kato2016} for the \SSS--\PPP transition, in~\cite{Franchi2017, Bouganne2017} for the \SSS--$^3$P$_0$ transition of Yb, and in~\cite{Campbell2017} for the \SSS--$^3$P$_0$ transition of strontium (Sr).
In contrast, as shown in Fig.~\ref{fig:projection} (a), a single, broad spectrum of coexisting SF and normal components was observed in the case of  a shallow optical lattice depth~\cite{Kato2016}.
The hopping time at small optical lattice depth (0.6 ms for 5$E_R$ depth) is comparable to the excitation time (0.5 ms in the case of Fig.~\ref{fig:projection}); therefore, the peaks of the observed spectra are not well separated.

A shorter excitation time is preferable for suppressing atom hopping during excitation. However, this causes spectral broadening of the resonance lines, significantly exceeding the separation between peaks under our conditions.
The frequency separation of the peaks is given by the collisional shift: $\Delta\nu_{\text{col}}=(U_{ge}-U_{gg})/h$.
Here, $U_{gg}(=U)$ is the on-site two-body interaction and 
$U_{ge}$ is the two-body interaction between the \SSS state ($\left| g \right>$) and \PPP state ($\left| e \right>$).

As an alternative, we have developed a new method: atom-number-projection spectroscopy.
In this approach, we increase the optical lattice depth quickly in order to freeze atom hopping, and then irradiate the atoms with an excitation light pulse.
The ramp-up time is $0.1$ ms from 5$E_R$ to 15$E_R$ and is faster than atom hopping, 
but sufficiently slow to prevent atom excitation into the higher band of the optical lattice ($\sim$ 20 kHz).
Figure~\ref{fig:projection}(b) shows spectra obtained using the atom-number-projection method, 
and the site-occupancy-resolved spectrum was indeed acquired for the shallow optical lattice.
Here, $n$-occupancies of up to five were distinguished with a separation of approximately $(U_{eg}-U_{gg})/h$.
The lower the resonance frequencies, the higher the occupation numbers {\it n} became.    
Note that our excellent resolution that allows different site occupancies to be distinguished is obtained 
by exploiting the optical transition between the \SSS and \PPP ($m_J=0$) electronic states of Yb atoms, 
which have quite different two-body interactions of $U_{eg}/h$ = -8.5 kHz and $U_{gg}/h$ = 3.2 kHz at 15$E_R$.
An additional advantage is that neither the \SSS nor \PPP ($m_J=0$) state is sensitive to a magnetic field, which enables acquisition of narrow spectra free from possible broadening due to magnetic field inhomogeneity.

\begin{figure}[tb]
	\includegraphics[width=9cm]{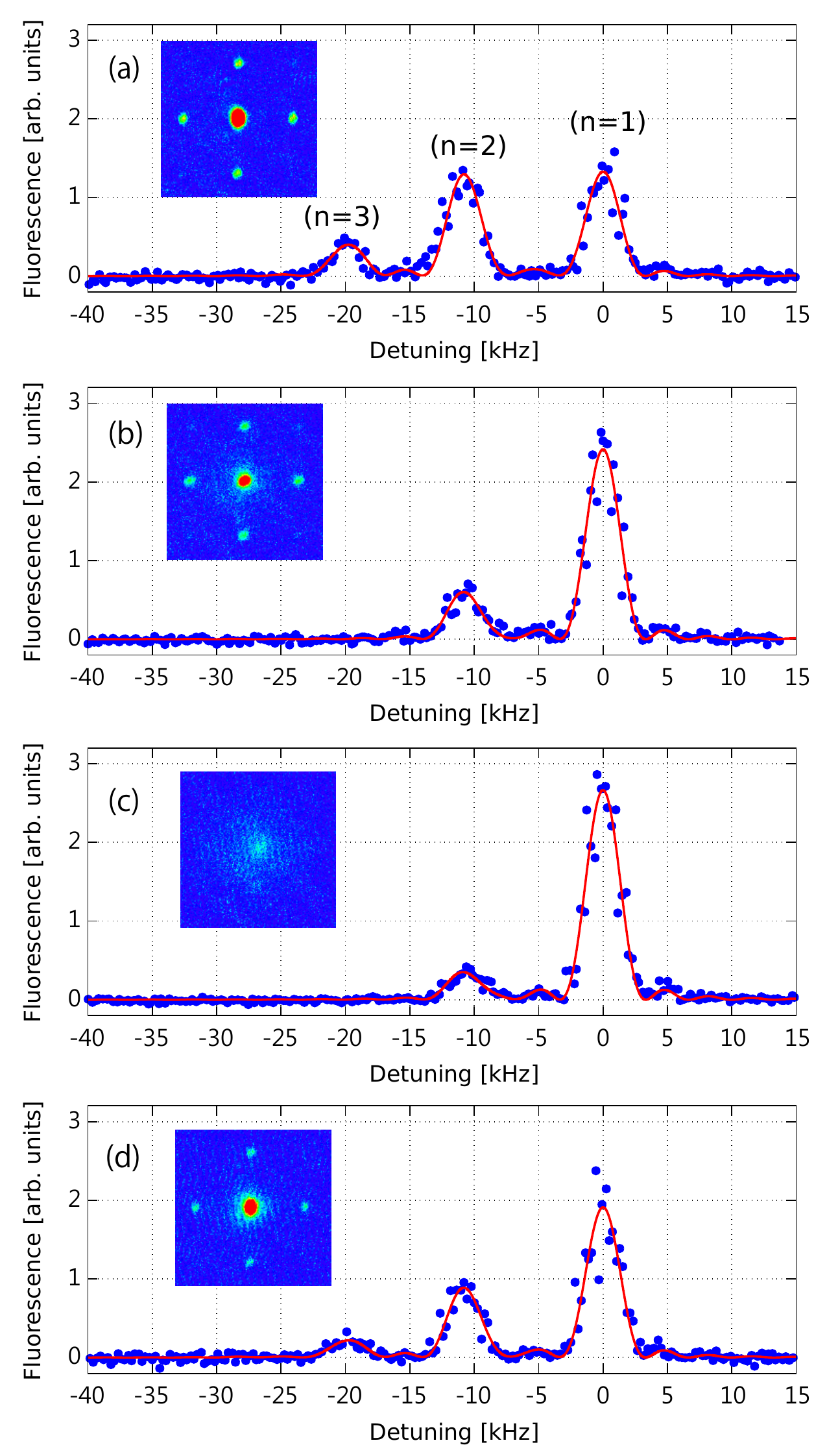}
	\caption{(color online) Typical high-resolution spectra obtained with atom-number-projection method. 
Three scans are superimposed after the long-term laser frequency drift is compensated. 
Insets: TOF images with the same experimental parameters. 
The TOFs are 14 ms. The total atom number is approximately 1.6 $\times 10^3$.
The solid red lines are fitting curves with a sinc function, with pulse width fixed at 0.3 ms. Lattice depths: (a)--(d) 5$E_R$, 10$E_R$, 15$E_R$, and 5$E_R$, respectively, and atomic entropies per atom: (a)--(d) 0.04, 0.05, 0.06, and 1.12 $k_b$, respectively. \label{fig:spectra}}
\end{figure}

Typical spectra are shown in Fig.~\ref{fig:spectra}, with the corresponding TOF images.
Naively, the area of each resonance in the spectrum is thought to be linearly proportional to the atom population in the corresponding occupancy of the optical lattice.
The excited state population $P_n(t_{\text{exc}})$ of the $n$-occupied site after the excitation time $t_{\text{exc}}$ is~\cite{Foot}
\begin{equation}
P_n(t_{\text{exc}})=\sin^2\left(\frac{\Omega_n t_{\text{exc}}}{2}\right)\exp\left(-t_{\text{exc}}\Gamma_n\right),
\end{equation}
where $\Omega_n$ are the (angular) Rabi frequencies and $\Gamma_n$ are the decay rates, with both parameters being dependent on $n$.
Note that resonance frequency shift $n\Delta\nu_{\text{col}}$ yields excitation of one atom only, even for an {\it n}-occupied site.
Thus, we must divide the $P_n(t_{\text{exc}})$ by $n$ when we consider the excited state population per atom.
If the spectral width of each resonance is the same, 
the area $A_n$ for the $n$-occupied site is linearly proportional to
\begin{equation}
N_nP_n(t_{\text{exc}})/n,
\end{equation}
where $N_n$ is the total atom number at the $n$-occupied site.

In the case of non-interacting atoms, the Rabi frequencies should be proportional to $\sqrt{n}$ because of the super-radiance or bosonic stimulation effect~\cite{Dicke1954, Gross1982}:
\begin{equation}
\Omega_n=\sqrt{n} \Omega_1.
\end{equation}
In addition, when the excitation time is much shorter; i.e., $\Omega_n t_{\text{exc}} \ll 1$ and $\Gamma_n t_{\text{exc}} \ll 1$, we obtain
\begin{equation}
A_n\propto N_n\frac{\Omega_1^2 t_{\text{exc}}^2}{4}.
\end{equation}
Because $\Omega_1$ and $t_{\text{exc}}$ were fixed in our experiment, the relative strengths of the areas indicate the relative atom number distributions among the sites in the optical lattice.

We experimentally examined the occupancy-dependent properties of the finite lifetime and transition probability in order to evaluate the total atom number at the $n$-occupied sites accurately.
We discuss these properties in the following subsections.
The details of the experimental parameters and procedures of our atom-number-projection spectroscopy are described in Appendix~\ref{ap:interaction}.


\subsection{Occupancy-dependent lifetime measurement}

\begin{figure}[tb]
	\includegraphics[width=9cm]{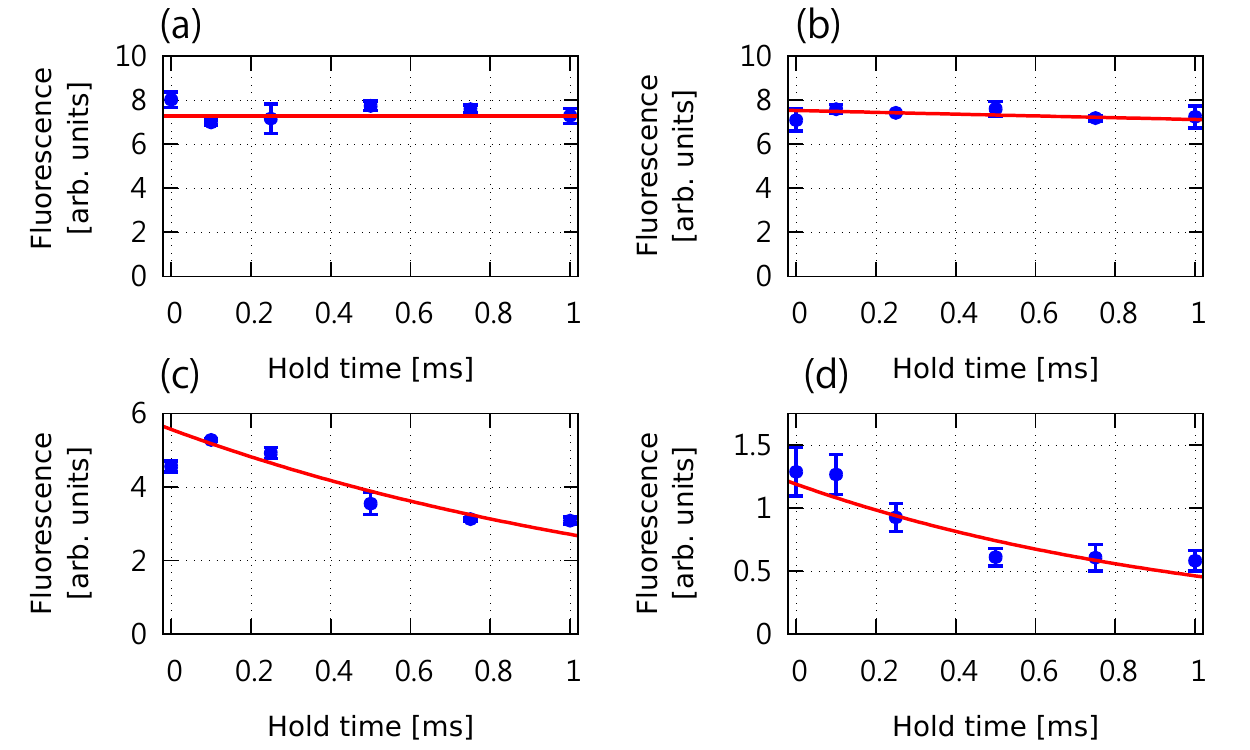}
	\caption{(color online) Lifetimes. The remaining excited-atom numbers as functions of hold time are measured for peaks in the $n$ = (a) $1$, (b) $2$, (c) $3$, and (d) $4$ occupied sites. The lifetimes of the $n$-occupied states ($n=2, 3, 4$) are $18(7)$, $1.4(2)$, and $1.1(3)$ ms, respectively. For the $n=1$ sites, we could not find any significant decay.\label{fig:lifetime}}
\end{figure}

The radiative lifetime of the \PPP state is approximately 15 s, 
which introduces negligible atom loss during our atom-number-projection spectroscopy.
Instead, the dominant loss process is
the inelastic collision between the atoms in the \SSS and \PPP states, 
which is induced by the fine-structure, principal-quantum-number, and Zeeman-state changing collisions~\cite{Uetake}.
Note that the magnetic sublevel \PPP ($m_J=0$) we use is not the lowest Zeeman-energy level.
Further, our excitation time of 0.3 ms is not negligible compared to the occupancy-dependent decay times, 
as shown in Fig.~\ref{fig:lifetime}; therefore, we actually measured the decay time to determine the correction factors 
for our atom-number-projection spectroscopy.

First, approximately $10^5$ BEC atoms were loaded into the shallow optical lattice (5$E_R$). 
Then, the lattice depth was suddenly increased to 15$E_R$ in 0.1 ms.
Next, we excited the atoms at the $n$-occupied sites 
and measured the number of atoms remaining in the \PPP state after a given hold time.
The results are shown in Fig.~\ref{fig:lifetime}.
We fit the data with single-exponential decay curves.
The measured decay constants of the $n$-occupied states $\tau_n$ ($n=2, 3, 4$) were $18(7)$, $1.4(2)$, and $1.1(3)$ ms, respectively.
For the $n=1$ site, we could not find any decay within our short hold time.
Therefore, it was necessary to correct the occupied atom number for the cases of $n=2$, $n=3$ and $n=4$ sites only, for which the correction factors were $1.02$, $1.24$, and $1.33$, respectively.

\subsection{Occupancy-dependent Rabi oscillation frequency}
For the non-interacting cases, the Rabi frequencies $\Omega_n/(2\pi)$ should be proportional to $\sqrt{n}$.
In the presence of the inter-atomic interaction, the situation is less simple and the $n$-dependence of the Rabi frequency is modified in general by broadening of the Wannier function due to inter-atom interactions ~\cite{Campbell2006, Franchi2017}. Such a modification was indeed observed in our system~\cite{Kato2016}. 
Here, we carefully evaluated the $n$-dependent Rabi frequency experimentally. 

\begin{figure}[tb]
	\includegraphics[width=9cm]{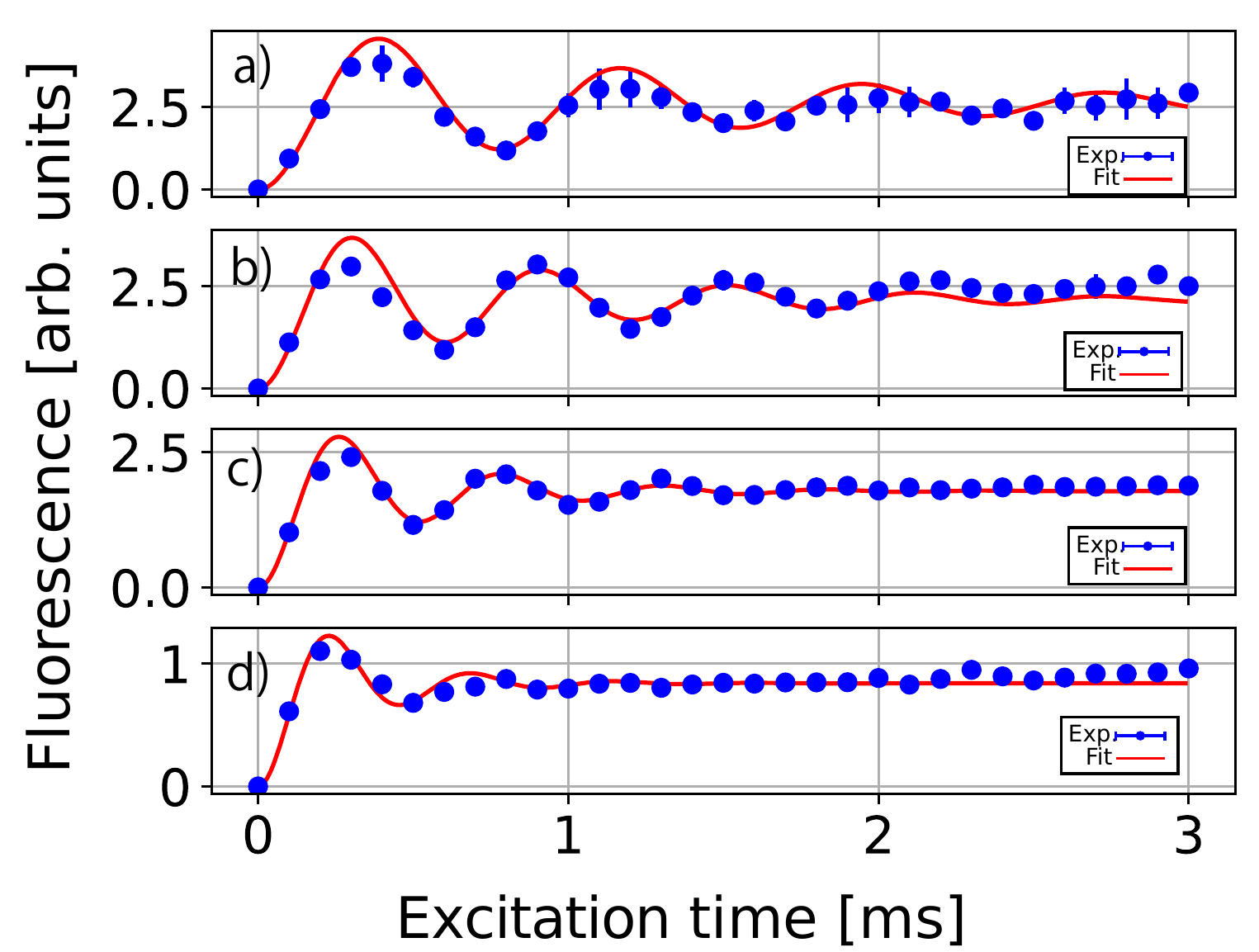}
	\caption{(color online) Rabi oscillations. 
Excited atom numbers as functions of excitation time measured by resonant peak excitation 
at $n=$ (a) $1$, (b) $2$, (c) $3$, and (d) $4$ occupied sites. 
The Rabi frequencies $\Omega_n/(2\pi)$ of the $n$-occupied sites are $1.28(1)$ ($n=1$), $1.65(3)$ ($n=2$), $1.92(3)$ ($n=3$), and $2.20(4)$ kHz ($n=4$).\label{fig:Rabi}}
\end{figure}

To observe clear Rabi oscillations, 
we excited the atoms with a relatively strong laser power of 4 mW, 
which corresponds to approximately 100 $\text{W}/\text{cm}^2$; the expected Rabi frequency at $n=1$ was approximately $2\pi\times2$ kHz.
The observed Rabi oscillations are shown in Fig.~\ref{fig:Rabi}.
The fitting lines were drawn by solving the optical Bloch equations numerically, assuming that the detuning was zero:
\begin{gather}
\frac{du_n}{dt}=-\frac{\Gamma_n}{2}u_n, \\
\frac{dv_n}{dt}=\Omega_n w_n - \frac{\Gamma_n}{2}v_n, \\
\frac{dw_n}{dt}=-\Omega_n v_n-\Gamma_n(w_n-1),
\end{gather}
where $u_n=\rho_{12,n}+\rho_{21,n}$, $v_n=-i(\rho_{12,n}-\rho_{21,n})$, $w_n=1-2\rho_{22,n}$, and $\rho_{ij,n}$ is the density matrix of the $n$-occupied sites.
The Rabi frequencies $\Omega_n/(2\pi)$ of the $n$-occupied sites were $1.28(1)$ ($n=1$), $1.65(3)$ ($n=2$), $1.92(3)$ ($n=3$), and $2.20(4)$ kHz ($n=4$).
The measured relative strength among the occupancy-dependent Rabi frequencies was used as a correction factor to estimate the atom-number distribution in our atom-number-projection spectroscopy.

\section{Possibility of Measuring Ensemble Average of Potential Energy Term} \label{sec:potential}


The third term of Eq. (\ref{eq:Hamiltonian}), the potential energy term, 
is from the inhomogeneous trap potential due to the FORT beams and optical lattice lasers.
Although the spatial distribution of the atoms in a trap for a single 2D plane can be directly measured using a high-spatial-resolution {\it in situ} imaging technique such as a quantum gas microscopy~\cite{Bakr2009, Sherson2010}, our imaging resolution was insufficient to accurately extract the spatial distributions of the atoms in our 3D optical lattice.

\section{Experimental Determination of Bose-Hubbard Energies} \label{sec:energy}
\begin{figure*}
	\includegraphics[width=17cm]{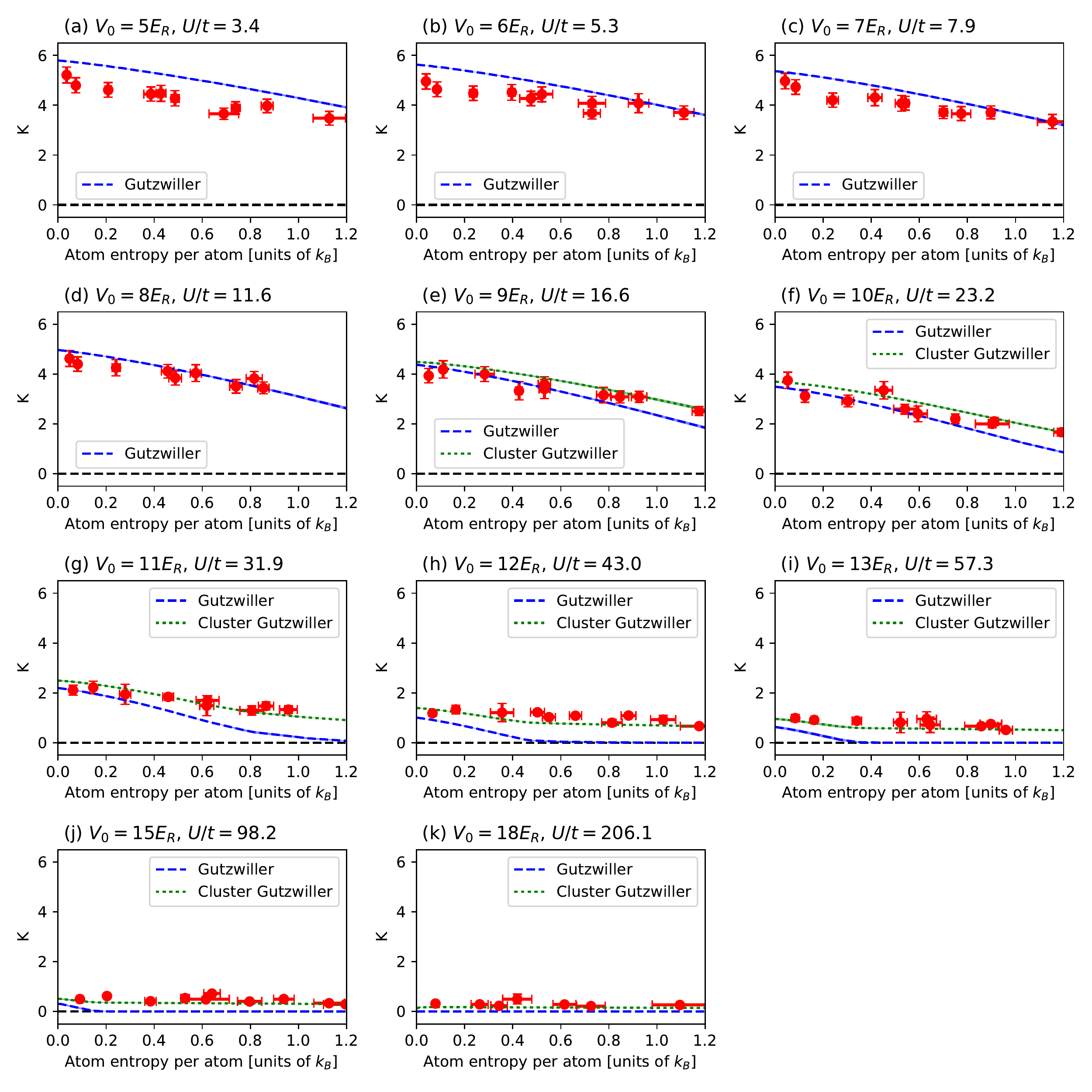}
	\caption{(color online) Measured ensemble averages of $\hat{K}=\sum_{\langle j, l \rangle}\left(\hat{a}^{\dagger}_j \hat{a}_l+h.c.\right)$ terms per atom as functions of atomic entropy per atom.
The dashed blue (dotted green) shadow lines indicate the results of numerical calculations based on the Gutzwiller (cluster-Gutzwiller) method. Different atom numbers ($1.4\times 10^3$ to $1.8\times10^3$) are represented by the shaded areas. 
The TOF images were taken immediately after atom loading onto the optical lattice.
For TOFs of 14 (18) ms, we took 10 (5) images and calculated the ensemble average for each one; 
then, these data were averaged.
Estimated difference between values before and after the correction of the finite-TOF effect is considered as systematic errors. The error bars indicate standard errors and include both systematic and statistical errors. See the text for details.\label{fig:K}}
\end{figure*}

Here, we present our main experimental results.
Figure~\ref{fig:K} shows the comprehensive measurements of the kinetic energy divided by the hopping matrix element $t$ per atom, i.e., the ensemble average of the term $\hat{K}=\sum_{\langle j, l \rangle}(\hat{a}^{\dagger}_j \hat{a}_l+h.c.)$ per atom for lattice depths from 5 to 18$E_R$ 
across the weakly to strongly interacting regimes as a function of the atomic entropy per atom.
When the lattice depth was 10.6$E_R$, $U/t$ was equal to $29.34$, which is the critical lattice depth for the SF-MI transition at $n=1$.
Note that the $\langle \hat{K}\rangle$ per atom must range between 6 and --6 for the 3D optical lattice.
The maximum and minimum values of 6 and --6 correspond to the atom condensation at $q=0$ and $q=\pm \pi/d_{lat}$, respectively.
Here, $q$ denotes the quasi-momentum.

Naturally, the expected $\langle \hat{K}\rangle$ behaviors were successfully observed in our experiment data, as shown in Fig.~\ref{fig:K}.
In a shallow optical lattice at sufficiently low entropy, almost all atoms should be condensed at $q=0$, corresponding to a $\langle \hat{K}\rangle$ close to 6; this is clearly apparent for the lower entropy data shown in Figs.~\ref{fig:K}(a)--(c).
With increased optical lattice depth, $\langle \hat{K}\rangle$ decreases and approaches zero because of the repulsive inter-atomic interaction ($U>0$); this is also clearly apparent as a general tendency of the data in Fig.~\ref{fig:K}.
In a deep optical lattice, all atoms are isolated and there is no phase coherence in a MI state. The atoms are distributed over the entire first Brillouin zone, 
corresponding to $\langle \hat{K}\rangle=0$; this behavior can be recognized in the data shown in Fig.~\ref{fig:K}(j) and (k).
When the atomic entropy increases, 
$\langle \hat{K}\rangle$ should decrease because of the thermal excitation to energetically higher states with larger $q$ at any lattice depth.
Again, this behavior can be clearly recognized as the general tendency of the data in Fig.~\ref{fig:K}.

\begin{figure*}
	\includegraphics[width=17cm]{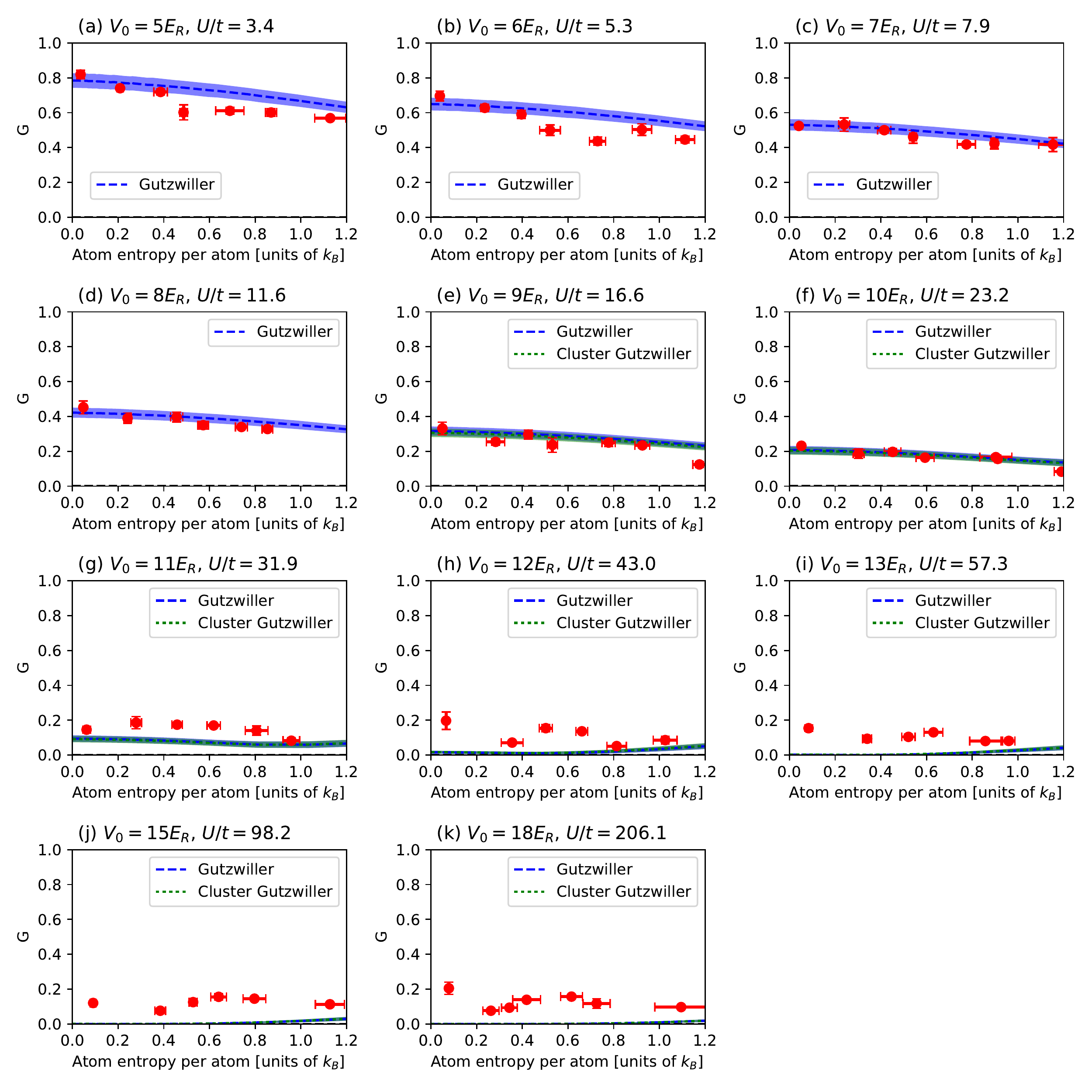}
	\caption{(color online) Measured ensemble averages of $\hat{G}=\sum_{j} \hat{a}^{\dagger}_j\hat{a}^{\dagger}_j\hat{a}_j\hat{a}_j$ per atom as functions of atomic entropy. The dashed blue (dotted green) shadow lines indicate the results of numerical calculations based on the Gutzwiller (cluster-Gutzwiller) method, with different atom numbers ($1.4\times 10^3$ to $1.8\times10^3$) being represented by shaded areas. The error bars indicate standard errors.\label{fig:g2}}
\end{figure*}

Figure~\ref{fig:g2} shows the comprehensive measurements of the interaction energy divided by $U/2$ per atom; namely, the ensemble average of the term
$\hat{G}=\sum_{j}\hat{a}^{\dagger}_j\hat{a}^{\dagger}_j\hat{a}_j\hat{a}_j$ per atom, 
again for lattice depths from 5 to 18$E_R$ 
across the weakly to strongly interacting regimes 
as a function of the atomic entropy per atom.
%

Again, the naturally expected behaviors of $\langle \hat{G}\rangle$ were successfully observed in our experiment data (Fig.~\ref{fig:g2}).
In a shallow optical lattice, the atom hopping process is dominant and the atoms are delocalized at multiply occupied sites, 
although the average filling-factor value is approximately unity.
This case yields a larger value of $\langle \hat{G}\rangle$, 
and this is clearly apparent in Figs.~\ref{fig:g2}(a)--(c), for example.
When the optical lattice depth is increased, 
the repulsive inter-atomic interaction plays a more important role 
in suppressing the atom hopping, 
yielding a decrease in $\langle \hat{G}\rangle$. This is clearly apparent as a general tendency of the data in Fig.~\ref{fig:g2}.
In a deep optical lattice, 
the atoms are isolated in an MI state with unit filling, 
which corresponds to $\langle \hat{G}\rangle=0$, as apparent in the data in Figs.~\ref{fig:g2}(h)--(k).
In the SF state, $\langle \hat{G}\rangle$ should decrease when the atomic entropy increases, 
because the thermal excitation yields expansion of the atomic cloud 
and a decrease in the multiply occupied sites.
This can be clearly recognized again as the general tendency of the data in Fig.~\ref{fig:g2}.

\begin{figure*}
	\includegraphics[width=17cm]{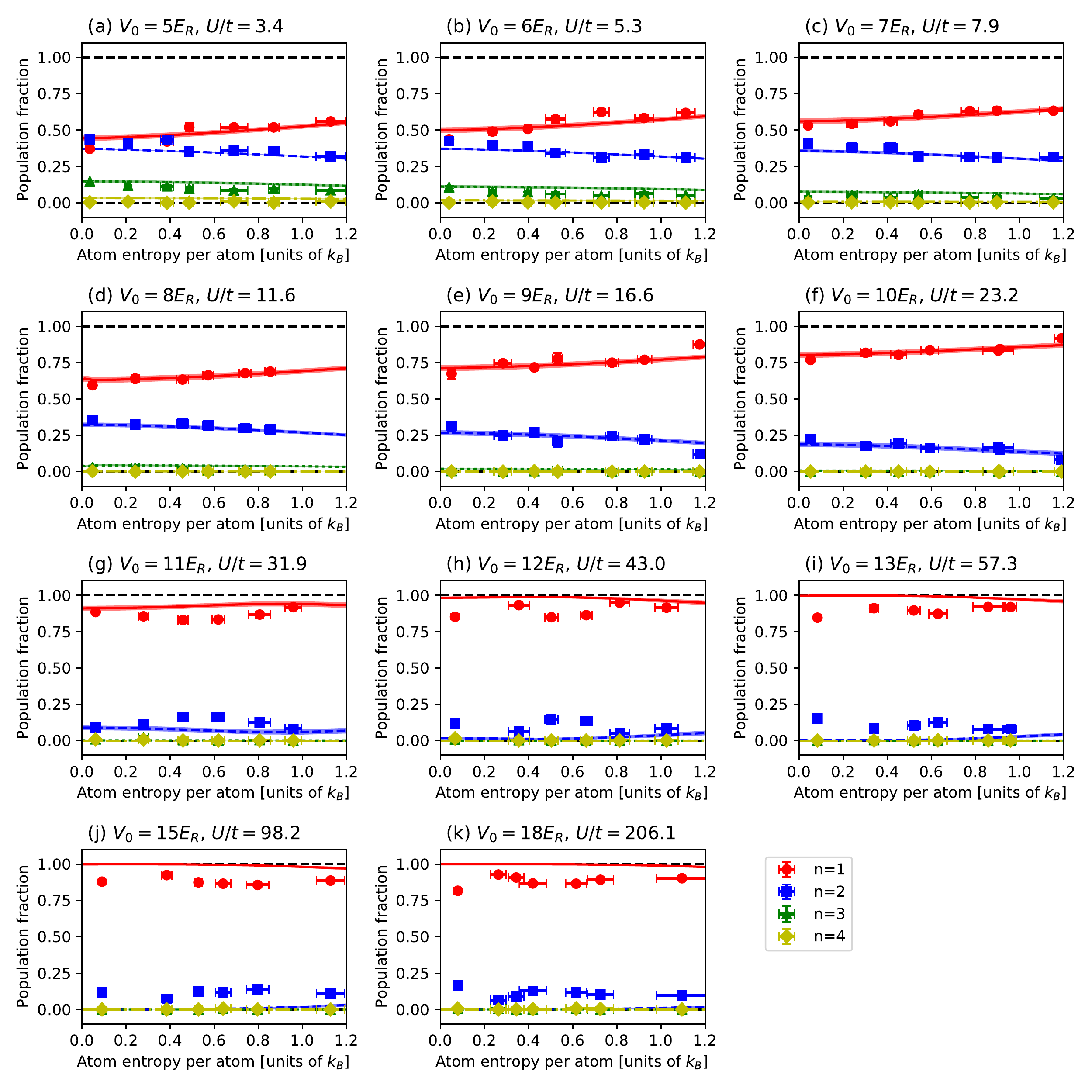}
	\caption{(color online) Population fractions as functions of atomic entropy per atom. The red circles, blue boxes, green triangles, and yellow diamonds show the normalized areas of $n=1$, $2$, $3$, and $4$ occupied sites, respectively. The solid red, dashed blue, dotted green, and dashed-dotted yellow lines indicate the numerical results for $n=1$, $2$, $3$, and $4$ occupied sites, respectively. The error bars show standard errors.\label{fig:weight}}
\end{figure*}

The population fractions, which could be directly measured by our atom-number-projection spectroscopy technique, 
elucidated further details of the atom number distribution in an optical lattice site. 
The population fractions at various lattice depths as functions of the atomic entropy are shown in Fig.~\ref{fig:weight}.
We found $n=3$ occupancy, although small, at small lattice depths only, as shown in Figs.~\ref{fig:weight}(a)--(c).
In Figs.~\ref{fig:weight}(a)--(g), decreases in the $n=2$ and $n=3$ populations accompanied by an increase in  the $n=1$ population can be clearly observed 
in accordance with the atomic entropy and lattice depth increase; this can be interpreted as originating from disappearance of the SF components.
 
%


\section{Numerical Calculation Benchmark} \label{sec:theory}

The obtained experimental results can be used as benchmarks for state-of-the-art numerical methods of quantum many-body theory.
As an illustrative example, in this section, we compare the measured kinetic and interaction energies as well as the population fractions of {\it n}-occupied sites with numerical calculations based on the Gutzwiller and cluster-Gutzwiller approximations.

\subsection{Gutzwiller approximation}
In this subsection, we explain two numerical methods based on Gutzwiller approximation.
One is a simple finite-temperature Gutzwiller approximation, where the effects of boson hopping are approximated as a mean field \cite{Kato2016,Sugawa2011}.
This is a simple local approximation obtained by solving the localized Hamiltonian with the exact diagonalization method at finite temperature.
Local thermodynamic quantities such as double occupancies can be well approximated by this calculation \cite{Kato2016,Sugawa2011}.
Another method is a cluster-type extension of this local approximation; that is, some of the hopping terms are included in the exact diagonalization calculation.
This cluster Gutzwiller approximation allows us to consider the kinetic energy much more effectively than the local approximation.


In a local Gutzwiller approximation, the Bose--Hubbard Hamiltonian ${\hat{\mathcal{H}}}$ is approximated by the set of effective local Hamiltonians
\begin{equation}
{\cal \hat{H}}_{{\rm loc},j}=F_{j}\hat{a}^\dag_{j}+F_{j}^*\hat{a}_{j}+(V_j-\mu)\hat{n}_{j}+U_{}\hat{a}_{j}^\dag\hat{a}_{j}^\dag\hat{a}_{j}\hat{a}_{j},
\end{equation}
under a self-consistent condition at thermal equilibrium for each local Hamiltonian. The mean field $F_{j}$ is given by
\begin{equation}
F_{j}=-\sum_l t_{jl}a_l,
\end{equation}
where $a_l=\langle \hat{a}_{l} \rangle$, $a_l^*=\langle \hat{a}_{l}^\dag \rangle$, and $t_{jl}=t$ for adjacent $j$ and $l$ sites, and $t_{jl}=0$ otherwise. The Hubbard parameters, including $t$, $U$, and $V_j$, are determined {\it ab initio} from Wannier functions.
We use the exact diagonalization method to solve these local $N_L$ Hamiltonians at finite temperature, where $N_L$ is the number of lattice sites. Here, we solve the finite Hilbert space by truncating states with a large number of bosons ($>8$) at each lattice site. The truncated states are negligible, because the on-site interaction suppresses them, even for shallow lattices. 

Under the self-consistent conditions, we calculate the double occupancy $\langle \hat{G}\rangle=\sum_j \langle \hat{a}_{j}^\dag\hat{a}_{j}^\dag \hat{a}_{j}\hat{a}_{j} \rangle$, potential energy $\sum_j V_j \langle \hat{a}_{j}^\dag \hat{a}_{j} \rangle$, and kinetic energy $-t\langle \hat{K}\rangle=-t\langle \hat{a}_{j}^\dag \rangle \langle \hat{a}_{l} \rangle$.
A non-local quantity such as $\langle \hat{a}_{j}^\dag \hat{a}_{l} \rangle$ is now approximated as a product of the local quantities $\langle \hat{a}_{j}^\dag \rangle \langle \hat{a}_{l} \rangle$.
The kinetic energy under the local approximation corresponds to the energy of the condensed bosons
\begin{equation}
-tK_{BEC}=-\sum_{jl} t_{jl} a_j^*a_l,
\end{equation} 
and the energies of the uncondensed normal states (normal fluid and MI) that appear as a result of the thermal fluctuation and correlation effects
\begin{equation}
-t\langle\hat{K}_{NS}\rangle=-\sum_{jl} t_{jl} \langle (\hat{a}_{j}^\dag-a^*_j)(\hat{a}_{l}-a_l )\rangle,
\end{equation}
are completely neglected.
Thus, the local Gutzwiller approximation inevitably underestimates the kinetic energies at middle-depth lattices.
In contrast, local quantities can be directly calculated using the exact-diagonalization method, which allows us to properly consider the effects of the normal states.


In a cluster-Gutzwiller approximation, the local Hamiltonians are extended to the two-site cluster Hamiltonians including a hopping term:
\begin{equation}
{\cal \hat{H}}_{{\rm TSC},jl}={\cal \hat{H}}_{{\rm loc},j}+{\cal \hat{H}}_{{\rm loc},l}-t_{jl}\hat{a}_j^\dag \hat{a}_l+h.c.
\end{equation}
We use exact diagonalization to solve the cluster Hamiltonian by truncating states with more than eight bosons in each cluster.
We also extend two self-consistency conditions in the cluster Hamiltonian ${\cal \hat{H}}_{{\rm TSC},jl}$:
\begin{eqnarray}
F_{j}&=&-\sum_{\alpha \ne l} t_{j\alpha}\langle \hat{a}_{\alpha} \rangle \\
F_{l}&=&-\sum_{\alpha \ne j} t_{l\alpha}\langle \hat{a}_{\alpha} \rangle.
\end{eqnarray}
That is, to avoid double counting of the effects of $t_{jl}\hat{a}_j^\dag \hat{a}_l+h.c.$, we subtract this term from the mean fields $F_j$ and $F_l$.
We solve $3N_L$ cluster Hamiltonians for the 3D cubic lattice, and local quantities such as $\langle \hat{a}_{j} \rangle$ are obtained from the average of six clusters ${\cal \hat{H}}_{{\rm TSC},j\alpha}$ for $\alpha \in$ sites adjacent to $j$.
Note that, when the self-consistency conditions are satisfied, the local quantities for the $j\,$th site in the ${\cal \hat{H}}_{{\rm TSC},j\alpha}$ agree well with each other. 
For ${\cal \hat{H}}_{{\rm TSC},jl}$, we can calculate a non-local quantity $\langle \hat{a}_{j}^\dag \hat{a}_{l} \rangle$, allowing us to obtain a kinetic energy that includes the effects of normal states, $-t\langle \hat{K}_{NS}\rangle$.

\subsection{Comparison of experiment and theory} \label{sec:discussion}
We compared the measured kinetic and interaction energies as well as the population fractions of {\it n}-occupied sites 
with numerical calculations of the Gutzwiller and cluster-Gutzwiller approximations in finite entropy (finite temperature).
Note that the trap potentials and particle numbers in the calculations were the same as those of the experiments and there were no fitting parameters in the calculation.

The dashed blue and dotted green lines in Fig.~\ref{fig:K}  represent the numerical results for the $\langle \hat{K}\rangle$ term obtained
using the Gutzwiller and cluster-Gutzwiller methods, respectively, with different atom numbers ($1.4\times 10^3$ to $1.8\times10^3$) being represented by the shaded areas.
Although we can observe overall agreement between the experimental data and numerical calculations for the overall lattice depth and atomic entropy, 
our measurement was highly consistent with the numerical calculation using the cluster-Gutzwiller method.

We discuss here possible origins of slight deference between the measured and numerical values appeared at the shallow optical lattice.
One of the possibility is imaging resolution of TOF images.
We consider the condensate atoms with zero momentum for simple explanation, which has a $\langle \hat{K}\rangle=6$ by definition. 
The finite imaging resolution broadens the measured momentum distribution around the zero momentum.
We consider a Gaussian point-spread function of $\left(1/\sqrt{2\pi}\Delta k\right)\exp\left(-k^2/(2\Delta k^2)\right)$ as the structure factor $S(k)$, where $\Delta k=m\sigma/(\hbar t_{\text{TOF}})$ and $\sigma$ is a resolution and assume $\Delta k d_{\text{lat}} \ll \pi$.
By simple calculation, the measured $\langle\hat{K}\rangle$ should be $K'=6\exp\left(-\Delta k^2 d^2_{\text{lat}}/2\right)$.
If $\sigma=$ 5 $\mu m$, $K'\sim$ 5.7 and therefore the finite resolution is not negligible when lattice depth is shallow and the system has large $\langle\hat{K}\rangle$.
Similar broadening might also occur when the interaction energy is transfered to the kinetic energy, which is discussed in Ref.~\cite{Kashurnikov2002}, where hydrodynamic expansion occurs around low-momentum part and results in peak broadening.

Both Gutzwiller methods exhibited similar results at shallow lattice depths, but differences emerged at deeper lattice depths and in the large-atomic-entropy regime.
The technical difference between the Gutzwiller and cluster-Gutzwiller methods lies in the handling of the atomic correlation of the nearest-neighbor sites.
In the case of the Gutzwiller method, the atomic correlation between the nearest-neighbor sites is from the SF component; thus, the atomic correlation of the thermal component is not considered.
Therefore, testing with our experimental data revealed that the atomic correlation between the nearest-neighbor sites from thermal fluctuation is indeed important in higher-entropy and deeper-lattice cases.

The dashed blue and dotted green lines in Fig.~\ref{fig:g2} represent the numerical results for the $\langle \hat{G}\rangle$ term obtained using the Gutzwiller and cluster-Gutzwiller methods, respectively,
with different atom numbers ($1.4\times 10^3$ to $1.8\times10^3$) being represented by the shaded areas.
In contrast to the $\langle \hat{K}\rangle$ term, the numerical results obtained using both the Gutzwiller and cluster-Gutzwiller methods were similar.
Again, overall agreement between the experimental data and numerical calculations was obtained for almost all lattice depths and atomic entropy.
However, differences between the measured and numerical values appeared when the optical lattice was deeper, as shown in Figs.~\ref{fig:g2}(g)--(k).
Although we are uncertain of the origin of these differences, we suspect that double occupancy may have occurred in the deeper optical lattice regime, because of the slight breaking of the adiabatic condition during lattice loading~\cite{Zakrzewski2009, Hung2010, Dolfi2015}.
Our ramp-up time of approximately 200 ms should be sufficient to reach local thermalization, 
but may be too short for global-mass redistribution in the deep-lattice case.
Note that the non-negligible atomic heating and loss observed for longer loading times limits us to this ramp-up time.

Numerical calculation of the population fractions as functions of the atomic entropy per atom was also performed, at various lattice depths.
The solid red, dashed blue, dotted green, and dashed-dotted yellow solid lines in Fig.~\ref{fig:weight} show the numerical results for the normalized areas of $n=1$, $2$, $3$, and $4$ occupied sites, respectively. 
Here, the total of the normalized areas is equal to unity.
We found excellent agreement between the experiment and numerical calculations, especially up to the critical lattice depth of 11$E_R$ (Figs.~\ref{fig:weight}(a)--(f)), but a certain disagreement at deeper lattice depth (Figs.~\ref{fig:weight}(g)--(k)), which can be attributed to the same reason discussed with regard to the disagreement for $\langle \hat{G}\rangle$ above.

\begin{figure*}
	\includegraphics[width=17cm]{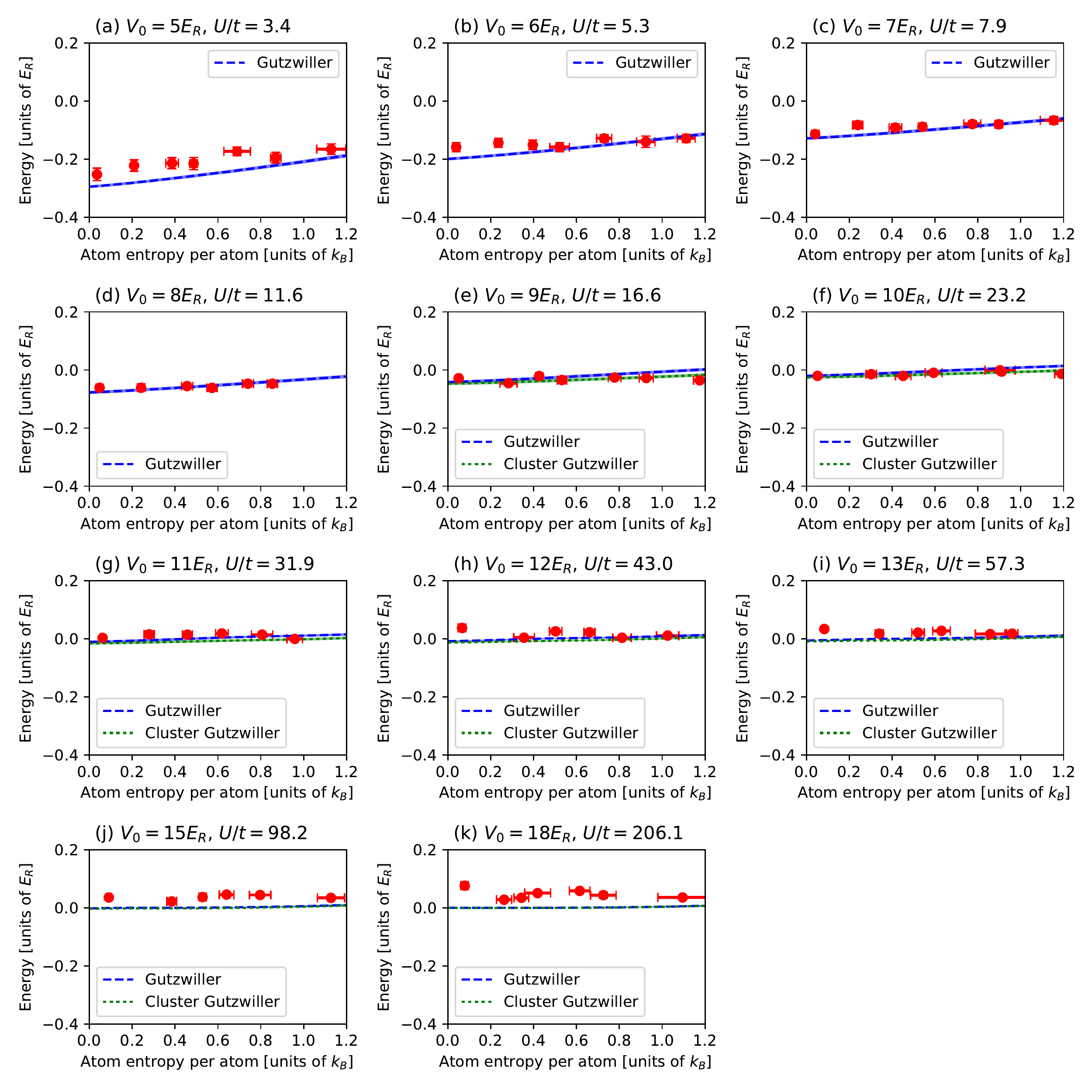}
	\caption{(color online) Total internal energies (i.e., the sum of the kinetic and interaction energies) per atom as functions of atomic entropy. The dashed blue (dotted green) shadow lines indicate the results of numerical calculations based on the Gutzwiller (cluster-Gutzwiller) method, with different atom numbers ($1.4\times 10^3$ to $1.8\times10^3$) being represented by the shaded areas. The error bars indicate standard errors.\label{fig:energy}}
\end{figure*}

We also investigated the total internal energy per atom (i.e., the sum of the kinetic and interaction energies) at various lattice depths as a function of atomic entropy (Fig.~\ref{fig:energy}).
%
The numerical results obtained using both Gutzwiller and cluster-Gutzwiller methods were similar.
The difference between the two numerical calculations of $\langle \hat{K}\rangle$ was not small at deeper lattice depth; however, the calculated kinetic energies of $-t\langle \hat{K}\rangle$ were almost identical because of the small values of $t$ at deeper lattice depth. 
The measured values were consistent with the numerical results.



\section{Conclusions and Future Prospects} \label{sec:conclusion}

We have presented, to our best knowledge, the first measurements of the ensemble averages of both the kinetic and interaction energies of the 3D Bose--Hubbard model at finite temperature and various optical lattice depths 
by establishing a protocol to accurately extract the ensemble average of the kinetic energy from a TOF signal and by developing a new method of atom-number-projection spectroscopy to accurately evaluate the interaction term across the weakly to strongly interacting regimes.
Our measurements showed rather strong dependence on the atomic entropy, except in the strongly correlated region.
This implies that information on the equilibrium state of the Bose--Hubbard system can be obtained from these measurements.
In addition, our atom-number-projection spectroscopy method offers information on the relative populations of the multiply occupied sites from the population fractions.
In this study, using these population fractions, we observed a decrease in the $n=2$ and $n=3$ populations when the atomic entropy and lattice depth increased; this behavior should be due to the disappearance of the SF components.
The obtained experimental results for the internal energies as well as the population fractions were compared with numerical calculations based on finite-temperature Gutzwiller and cluster-Gutzwiller methods; hence, we obtained agreement between the experiment and cluster-Gutzwiller calculation without fitting parameters. This indicates the important role of the atomic correlation between the nearest-neighbor sites through thermal fluctuation, especially in higher-entropy and deeper-lattice cases.

Measurement of the internal energy for various entropies offers a novel possibility of estimating the atomic temperature in a lattice, which is the most important parameter governing the thermal equilibrium state.
If the total internal energies, i.e., the kinetic, interaction, and potential terms, are measured experimentally, one can determine the temperature $T$ using the thermodynamic relation $T=\partial E/\partial S$,
where $E$ is the total internal energy and $S$ is the atomic entropy.
We have checked this proposal numerically (see Appendix~\ref{ap:temperature}).
This possibility is important, because the temperature in an optical lattice has only been estimated indirectly to date, through comparison of the experimental results and theoretical calculation.
Finally, the methods demonstrated here are not particular to Bose gases in equilibrium, but can be applied to Fermi gases, Bose--Fermi mixtures, and even non-equilibrium states.

This paper is, to the best of our knowledge, the first report of experimental determination of both the kinetic and interaction energies of quantum many-body systems.
This study offers a unique advantage of cold atom system for ``quantum simulators''.


\acknowledgements{We thank H. Shiotsu, K. Takiguchi, and J. Sakamoto for  experimental assistance.
	This work is supported by MEXT/JSPS KAKENHI, Grant Numbers JP25220711, JP26247064, JP16H00990, JP16H01053, JP16H00801, JP18H05228, JP18H05405;
	and the Impulsing Paradigm Change through Disruptive Technologies (ImPACT) program; and CREST, JST JPMJCR 1673; and the Matsuo Foundation.}

Y. Takasu and Y.N. equally contributed to this work.


\appendix
\section{Energy diagram and scattering length of Yb}\label{ap:energy}
\begin{figure}[tb]
	\includegraphics[width=9cm]{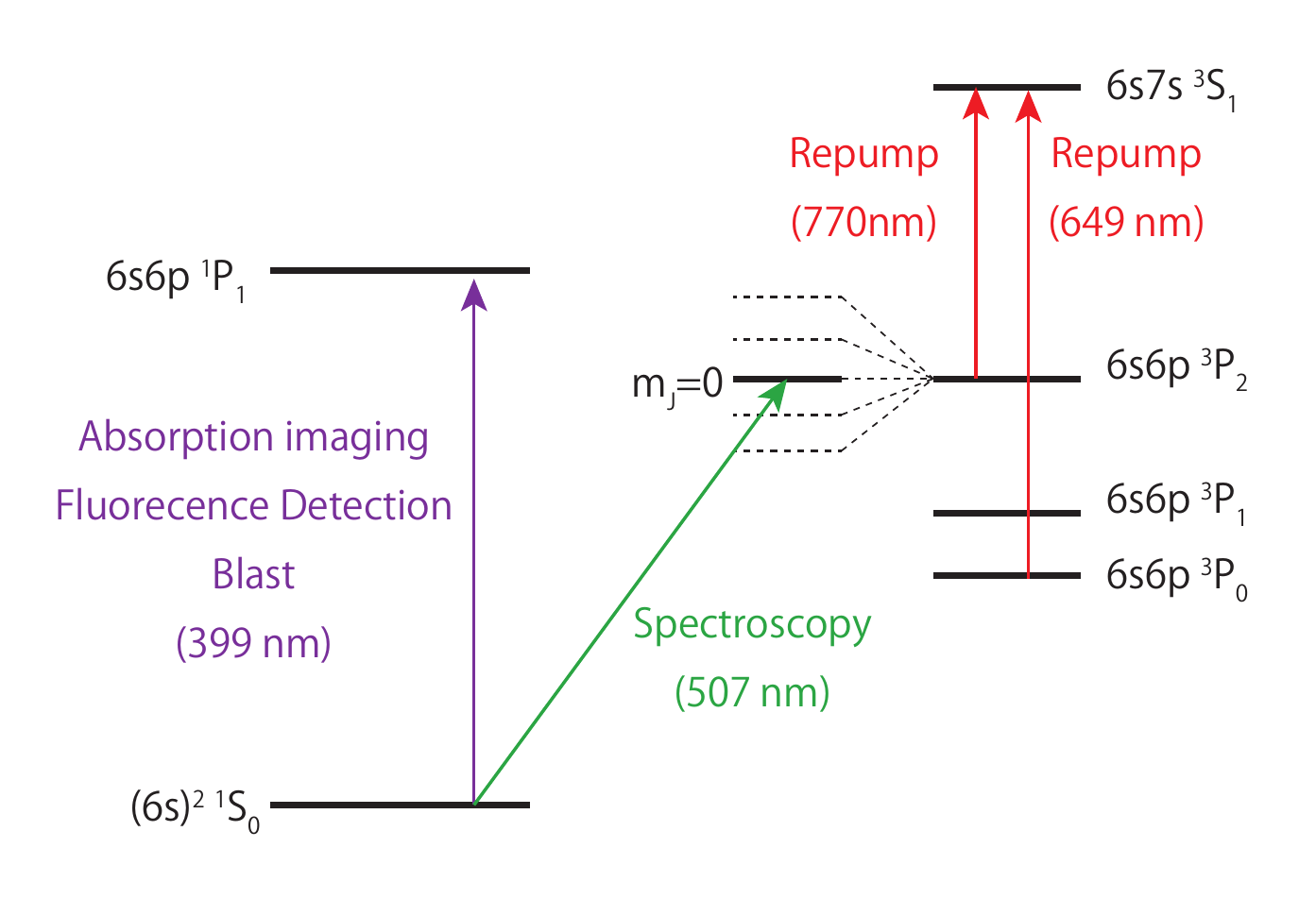}
	\caption{(color online) Schematic energy diagram (not scaled) of Yb relevant to the experiment.\label{fig:energydiagram}}
\end{figure}

Figure~\ref{fig:energydiagram} shows the Yb schematic energy diagram (not scaled) relevant to the experiment.
Throughout this paper, we used the value of the scattering length of ${}^{174}\text{Yb}$ of 5.55 nm~\cite{Kitagawa2008}.

\section{Additional information of our Experimental Setup and Procedure} \label{ap:setup}
The beam waists ($1/e^2$ radii) of the horizontal FORT were approximately 15 and 33 $\mu$m and the short axes of the ellipses were oriented along the Z-axis.
The beam waists of the vertical FORT were approximately 43 and 126 $\mu$m, and the short axes of the ellipses were oriented along the X'-axis, where the X'-axis formed an angle of 45 degrees relative to both the X- and Y-axes.
The beam waists of the lattice beams were approximately 100 $\mu$m.
The FORT trap frequencies were (27.9, 130, 162.5) Hz after the lattice loading.

\begin{figure}[tb]
	\includegraphics[width=9cm]{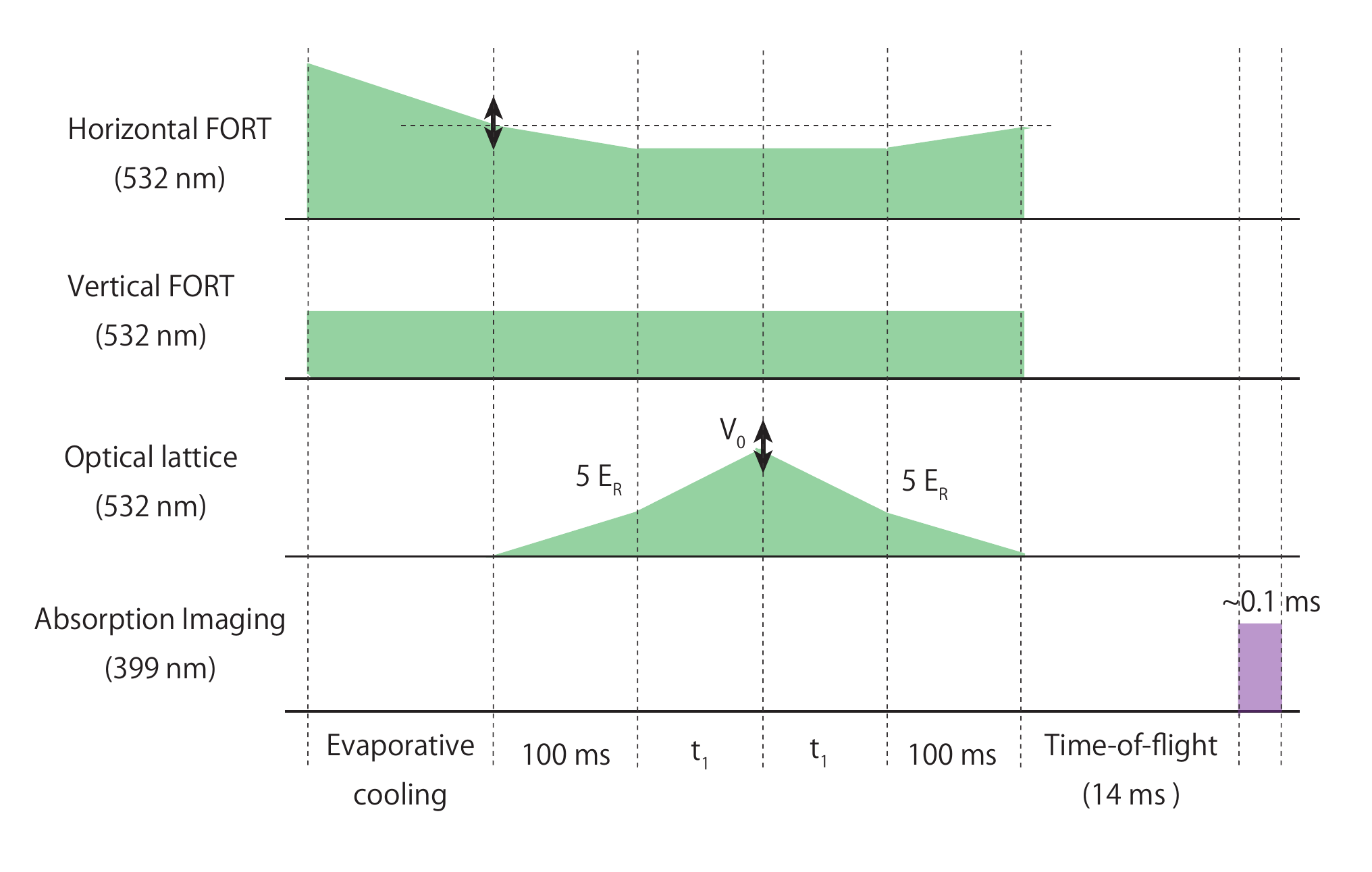}
	\caption{(color online) Schematic time sequence (not scaled) for entropy measurements. Here, $t_1=10(V_0/E_R-5)$ [ms]. The double-sided arrows indicate variable parameters (see text for details).\label{fig:sequence-entorpy}}
\end{figure}

The measurement procedure for the entropy and atom number after adiabatically ramping down the optical lattice in reverse order is shown in Fig.~\ref{fig:sequence-entorpy}.

The optical lattice depth is calibrated by a pulsed optical lattice method (see also~\cite{Kato2016, Denschlag2002}).

\section{Kinetic term} \label{ap:TOF}
In this section, we label the atom momentum $\mathbf{k}_{\text{TOF}}$ as $\mathbf{k}$ for simplicity. 
The atomic density distribution after the TOF $t_{\text{TOF}}$, i.e., $n(\mathbf{k}, \tTOF)$, is
\begin{equation}
	n(\mathbf{k}, \tTOF)=\left(\frac{m}{\hbar t_{\text{TOF}}}\right)^3 \left|\tilde{w}_0(\mathbf{k})\right|^2 S(\mathbf{k},\tTOF).
\end{equation}
The atomic momentum after the TOF is calculated from the positions of the atoms $\mathbf{r}_{\text{TOF}}$ and $\mathbf{k}$ = $m\mathbf{r}_{\text{TOF}}/\hbar t_{\text{TOF}}$.
Here, $\tilde{w}_0 (\mathbf{k})$ is the Fourier transformation of the Wannier function in the lowest Bloch band $w_0(\mathbf{r})$ and
\begin{align}
&\tilde{w}_0 (\mathbf{k})=\iiint w_0(\mathbf{r}) e^{i\mathbf{k}\cdot\mathbf{r}} d\mathbf{r} \nonumber \\
&=\int w_0(x)e^{ixk_x} dx\int w_0(y)e^{iyk_y}dy\int w_0(z)e^{izk_z}dz.
\end{align}
The structure factor $S(\mathbf{k})$ is
\begin{equation}
	S(\mathbf{k},\tTOF)=\sum_{j,l}e^{i \mathbf{k}\cdot \left(\mathbf{r}_j-\mathbf{r}_l\right)-i\left(\frac{m}{2\hbar t_{\text{TOF}}}\right)(\mathbf{r}^2_j-\mathbf{r}^2_l)} \langle \hat{a}^{\dagger}_j \hat{a}_l \rangle.
\end{equation}

First, we consider the integral of $n(\mathbf{k}=m\mathbf{r}_{\text{TOF}}/\hbar t_{\text{TOF}}, \tTOF)$ along the Y-axis in position space:
\begin{align}
&\int n\left(\mathbf{k}=\frac{m\mathbf{r}_{\text{TOF}}}{\hbar t_{\text{TOF}}}, \tTOF\right) dy_{\text{TOF}}=\frac{\hbar\tTOF}{m}\int n(\mathbf{k}, \tTOF)dk_y \nonumber\\
&=A^2\int dk_y|\tilde{w}_0(\mathbf{k})|^2\sum_{j,l}e^{i \mathbf{k}\cdot \left(\mathbf{r}_j-\mathbf{r}_l\right)-i\left(\frac{m}{2\hbar t_{\text{TOF}}}\right)(\mathbf{r}^2_j-\mathbf{r}^2_l)} \langle \hat{a}^{\dagger}_j \hat{a}_l \rangle \nonumber\\
&=A^2|\tilde{w}_0(\mathbf{k}_{\perp})|^2\sum_{j,l} \langle \hat{a}^{\dagger}_j \hat{a}_l \rangle e^{i\mathbf{k}_{\perp}\cdot(\mathbf{r}_{\perp,j}-\mathbf{r}_{\perp,l})-i\left(\frac{m}{2\hbar t_{\text{TOF}}}\right)(\mathbf{r}^2_j-\mathbf{r}^2_l)} \nonumber \\
&\qquad\times\int dk_y |\tilde{w}_0(k_y)|^2e^{ik_y(y_j-y_l)}, \label{eq:Kia1}
\end{align}
where $A=m/(\hbar \tTOF)$, $\mathbf{r}_{\perp}=(x, z)$, and $\mathbf{k}_{\perp}=(k_x, k_z)$. Then,
\begin{align}
&\int dk_y |\tilde{w}_0(k_y)|^2e^{ik_y(y_j-y_l)} \nonumber\\
&=\int dk_y\int du' w_0^{*}(u')e^{-ik_y u'} \int du w_0(u)e^{ik_yu}e^{ik_y(y_j-y_l)} \nonumber \\
&=\int du\int du'w_0^{*}(u')w_0(u)\int dk_y e^{ik_y(u-u'+y_j-y_l)} \nonumber \\
&=\int du\int du'w_0^{*}(u')w_0(u)\delta(u-u'+y_j-y_l)\nonumber \\
&=\int w_0^{*}(u+y_j-y_l)w_0(u) du\nonumber \\
&=\begin{cases}
 1 & (y_j=y_l)\\
 0 & \text{otherwise}
\end{cases},\label{eq:Kia2}
\end{align}
where we use the orthogonality of the Wannier functions,
\begin{equation}
\int w_0^{*}(u+nd_{\text{lat}})w_0(u)du=
\begin{cases}
 1 & (n=0)\\
0 & \text{otherwise}
\end{cases},
\end{equation}
with $n$ = 0, $\pm1$, $\pm2$, $\cdots$, and $d_{\text{lat}}$ being the lattice spacing.

By applying Eq.~(\ref{eq:Kia2}) to Eq.~(\ref{eq:Kia1}), we obtain
\begin{align}
&\int n\left(\mathbf{k}=\frac{m\mathbf{r}_{\text{TOF}}}{\hbar t_{\text{TOF}}}, \tTOF\right) dy_{\text{TOF}} \nonumber\\
&=A^2|\tilde{w}_0(\mathbf{k}_{\perp})|^2\sum_{j,l} \langle \hat{a}^{\dagger}_j \hat{a}_l \rangle e^{i\mathbf{k}_{\perp}\cdot(\mathbf{r}_{\perp,j}-\mathbf{r}_{\perp,l})-i\left(\frac{m}{2\hbar t_{\text{TOF}}}\right)(\mathbf{r}^2_{\perp,j}-\mathbf{r}^2_{\perp,l})} \nonumber \\
&\qquad\times\delta_{y_j,y_l},
\end{align}
where $\delta_{x,y}$ is the Kronecker delta. That is, $\delta_{x,y}$=1 if and only if $x=y$; otherwise, $\delta_{x,y}=0$.

Similarly, we obtain the linear atomic density $n(k_x)$ is 
\begin{align}
&\iint n\left(\mathbf{k}=\frac{m\mathbf{r}_{\text{TOF}}}{\hbar t_{\text{TOF}}}, \tTOF\right) dy_{\text{TOF}}dz_{\text{TOF}}  \nonumber\\
&=A|\tilde{w}_0(k_x)|^2\sum_{j,l} \langle \hat{a}^{\dagger}_j \hat{a}_l \rangle e^{ik_x(x_j-x_l)-i\left(\frac{m}{2\hbar t_{\text{TOF}}}\right)(x_j^2-x_l^2)} \nonumber \\
&\qquad\times\delta_{y_j,y_l}\delta_{z_j,z_l} \label{eq:Kia3}
\end{align}
and
\begin{align}
&\iiint n\left(\mathbf{k}=\frac{m\mathbf{r}_{\text{TOF}}}{\hbar t_{\text{TOF}}}, \tTOF\right) dx_{\text{TOF}}dy_{\text{TOF}}dz_{\text{TOF}}  \nonumber\\
&=\sum_{j,l} \langle \hat{a}^{\dagger}_j \hat{a}_l \rangle \delta_{x_j,x_l}\delta_{y_j,y_l}\delta_{z_j,z_l} \nonumber \\
&=\sum_{j}\langle \hat{a}^{\dagger}_j \hat{a}_j \rangle=N,
\end{align}
where $N$ is the total number of atoms.

\subsection{Case I: Infinite TOF}

First, for simplicity, we consider the case in which the TOF is infinite and the Fresnel term $\exp[-im(\mathbf{r}_j^2-\mathbf{r}_l^2)/(2\hbar t_{\text{TOF}})]$ is negligible.
In this case, the linear atomic density $n(k_x)$ is (see also Eq.~(\ref{eq:Kia3})) as follows:
\begin{align}
n(k_x) &=A|\tilde{w}_0(k_x)|^2\sum_{j,l} \langle \hat{a}^{\dagger}_j \hat{a}_l \rangle e^{ik_x(x_j-x_l)} \delta_{y_j,y_l}\delta_{z_j,z_l}. \label{eq:Ki3}
\end{align}
Therefore, the ensemble average of the atomic correlations of nearest-neighbor sites, $\sum_{\langle j, l\rangle} \langle\hat{a}^\dagger_j\hat{a}_l\rangle$, is
obtained using Fourier transformation in the first Brillouin zone, where
\begin{align}
&\frac{d_{\text{lat}}}{2\pi}\int_{-\pi/d_{\text{lat}}}^{\pi/d_{\text{lat}}}\frac{n(k_x)}{A|\tilde{w}_0(k_x)|^2} e^{id_{\text{lat}}k_x} dk_x \nonumber \\
&=\frac{d_{\text{lat}}}{2\pi}\sum_{j,l} \langle\hat{a}^\dagger_j\hat{a}_l\rangle \delta_{y_j,y_l}\delta_{z_j,z_l} \int _{-\pi/d_{\text{lat}}}^{\pi/d_{\text{lat}}}e^{ik_x(x_j-x_l+d_{\text{lat}})} dk_x\nonumber\\
&=\sum_{j,l} \langle\hat{a}^\dagger_j\hat{a}_l\rangle \delta_{x_j,x_l-d_{\text{lat}}}\delta_{y_j,y_l}\delta_{z_j,z_l} \nonumber \\
&=\sum_j \langle\hat{a}^\dagger_j\hat{a}_{j+1}\rangle.
\end{align}

Similarly,
\begin{align}
&\frac{d_{\text{lat}}}{2\pi}\int_{-\pi/d_{\text{lat}}}^{\pi/d_{\text{lat}}}\frac{n(k_x)}{A|\tilde{w}_0(k_x)|^2} e^{-id_{\text{lat}}k_x} dk_x \nonumber \\
&=\sum_{j,l} \langle\hat{a}^\dagger_j\hat{a}_l\rangle \delta_{y_j,y_l}\delta_{z_j,z_l}\delta_{x_j,x_l-d_{\text{lat}}}.
\end{align}

Noted that if TOF images are symmetric with respect to the $k=0$, $\langle \hat{a}^\dagger_j\hat{a}_{l}\rangle=\langle \hat{a}^\dagger_l\hat{a}_{j}\rangle$ and therefore $\sum_{\langle j, l\rangle}\langle \hat{a}^\dagger_j\hat{a}_{l}\rangle$ is real.
This is valid if the hopping matrix element $t$ is real and the system is in equilibrium states (strictly speaking, if the system has time-reversal symmetry), because the kinetic energy $-t\sum_{\langle j, l\rangle}\langle \hat{a}^\dagger_j\hat{a}_{l}\rangle$ itself is required to be real.
This assumption is invalid for some special cases, for example, non-equilibrium states with non-zero total quasi-momentum, and equilibrium states with an artificial gauge field (complex hopping matrix elements)~\cite{Dalibard2011}.
In these cases, however, $\sum_{\langle j, l\rangle}\langle \hat{a}^\dagger_j\hat{a}_{l}\rangle$ must be complex and we believe that the kinetic energy can be obtained using the above procedure, if the Wannier functions are well defined.

Similarly,
\begin{align}
&\frac{d_{\text{lat}}}{2\pi}\int_{-\pi/d_{\text{lat}}}^{\pi/d_{\text{lat}}} \frac{n(k_x)}{A|\tilde{w}_0(k_x)|^2} dk_x \nonumber \\
&=\frac{d_{\text{lat}}}{2\pi}\sum_{j,l} \langle\hat{a}^\dagger_j\hat{a}_l\rangle \delta_{y_j,y_l}\delta_{z_j,z_l} \int_{-\pi/d_{\text{lat}}}^{\pi/d_{\text{lat}}} e^{ik_x(x_j-x_l)} dk_x\nonumber\\
&=\sum_{j,l} \langle\hat{a}^\dagger_j\hat{a}_l\rangle \delta_{x_j,x_l}\delta_{y_j,y_l}\delta_{z_j,z_l} \nonumber \\
&=\sum_{j}\langle\hat{a}^\dagger_j\hat{a}_j\rangle =N.
\end{align}

\subsection{Case II: Finite TOF}

\begin{figure}[tb]
	\includegraphics[width=9cm]{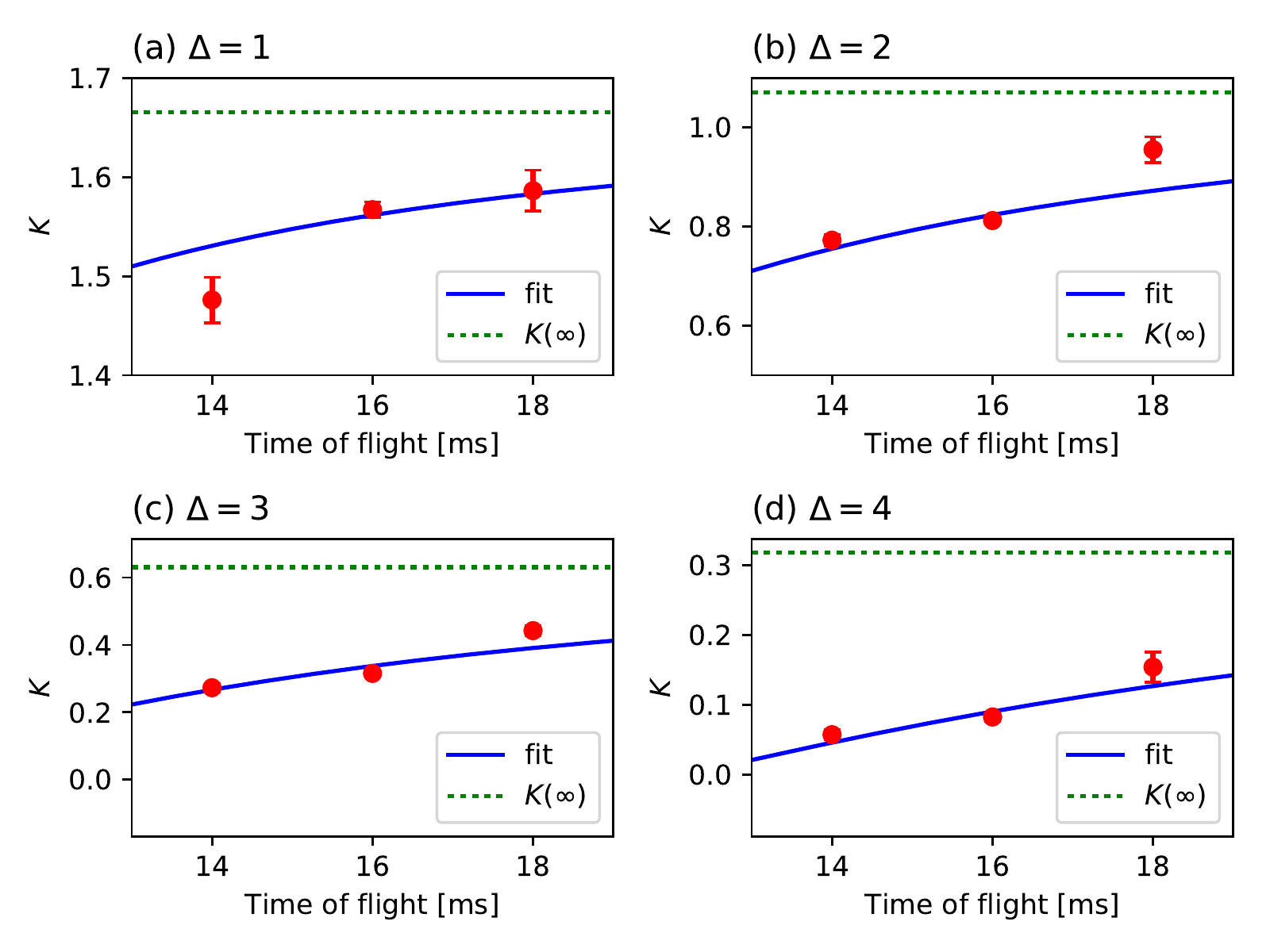}
	\caption{(color online) Measured long-range atomic correlations $\sum\langle\hat{a}^\dagger_j\hat{a}_{l}\rangle$ with lattice separations $\Delta l$ and their finite TOF corrections. 
The solid blue and dotted green lines indicate the fitting Eq. (\ref{eq:correction}) and the corrected values at infinite TOF, respectively. 
Because the data shown in this graph were used for the check of our compensation method, 
the atom number was $5\times 10^4$, different from the experimental value shown in the main text. 
The lattice depth was $V_0= 5E_R$. 
$\Delta l$: (a)1, (b) 2, (c) 3, and (d) 4. \label{fig:correction}}
\end{figure}

Under our experimental conditions, the Fresnel term is non-negligible because there is a finite TOF.
However, the effect is small; therefore, we can consider it to be a correction factor:
\begin{align}
&\frac{d_{\text{lat}}}{2\pi}\int_{-\pi/d_{\text{lat}}}^{\pi/d_{\text{lat}}} \frac{n(k_x)}{A|\tilde{w}_0(k_x)|^2} e^{id_{\text{lat}}k_x} dk_x \nonumber \\
&=\frac{d_{\text{lat}}}{2\pi}\sum_{j,l} \langle\hat{a}^\dagger_j\hat{a}_l\rangle \delta_{y_j,y_l}\delta_{z_j,z_l} e^{-i\left(\frac{m}{2\hbar\tTOF}\right)(x_j^2-x_l^2)}\nonumber \\
&\qquad\times\int _{-\pi/d_{\text{lat}}}^{\pi/d_{\text{lat}}}e^{ik_x(x_j-x_l+d_{\text{lat}})} dk_x\nonumber\\
&\sum_{j,l} \langle\hat{a}^\dagger_j\hat{a}_l\rangle \delta_{y_j,y_l}\delta_{z_j,z_l} \delta_{x_j,x_l-d_{\text{lat}}} e^{-i\left(\frac{m}{2\hbar\tTOF}\right)(x_j^2-x_l^2)} \nonumber \\
&=\sum_{j=1}^{N_L}\langle\hat{a}^\dagger_j\hat{a}_{j+1}\rangle e^{i\left(\frac{m}{2\hbar\tTOF}\right)(d_{\text{lat}}^2+2d_{\text{lat}}x_j)}.
\end{align}
Now, we assume that $\langle\hat{a}^\dagger_j\hat{a}_{j+1}\rangle$ is independent of site index $j$ and have a average value $\overline{\langle\hat{a}^\dagger_j\hat{a}_{j+1}\rangle}$, and the summation is from $x_1=-N'_Ld_{\text{lat}}$ to $x_{N_L-1}=(N'_{L}-1)d_{\text{lat}}$ and $N_L=2N'_L+1$,
\begin{align}
(\text{C14})\sim&\overline{\langle\hat{a}^\dagger_j\hat{a}_{j+1}\rangle}\sum_{j=1}^{N_L-1}~e^{i\left(\frac{m}{2\hbar\tTOF}\right)(d_{\text{lat}}^2+2d_{\text{lat}}x_j)} \nonumber \\
&=\overline{\langle\hat{a}^\dagger_j\hat{a}_{j+1}\rangle}~e^{i\left(\frac{md_{\text{lat}}^2}{2\hbar\tTOF}\right)} \sum_{j=-N'_L}^{N'_L-1}e^{i\left(\frac{md_{\text{lat}}^2j}{\hbar\tTOF}\right)} \nonumber \\
&=\overline{\langle\hat{a}^\dagger_j\hat{a}_{j+1}\rangle}e^{\left(iz/2-izN'_L\right)}\frac{1-e^{2iN'_Lz}}{1-e^{iz}}.
\end{align}
where $z=\frac{md_{\text{lat}}^2}{\hbar\tTOF}$.
We here defined the correction factor $C_1$ as
\begin{equation}
\frac{1}{C_1}=\frac{1}{2N'_L} \text{Re}\left[e^{\left(iz/2-izN'_L\right)}\frac{1-e^{2iN'_Lz}}{1-e^{iz}}\right],
\end{equation}
where we assume that $\langle\hat{a}^\dagger_j\hat{a}_{j+1}\rangle$ is real and therefore the correction factor is also real. 
Similarly, we obtain the relation on the long-range atomic correlation $\sum\langle\hat{a}^\dagger_j\hat{a}_{j+\Delta l}\rangle$ with the lattice separation $|\mathbf{r}_j-\mathbf{r}_l|=d_{\text{lat}}\Delta l$,
\begin{align}
&\frac{d_{\text{lat}}}{2\pi}\int_{-\pi/d_{\text{lat}}}^{\pi/d_{\text{lat}}} \frac{n(k_x)}{A|\tilde{w}_0(k_x)|^2} e^{i\Delta l d_{\text{lat}}k_x} dk_x \nonumber \\
&\sim \overline{\langle\hat{a}^\dagger_j\hat{a}_{j+\Delta l}\rangle}e^{i\left(\frac{md_{\text{lat}}^2\Delta l^2}{2\hbar\tTOF}\right)} \sum_{j=-N'_L}^{N'_L-\Delta l}e^{i\left(\frac{md_{\text{lat}}^2j\Delta l}{\hbar\tTOF}\right)} \nonumber \\
&=\overline{\langle\hat{a}^\dagger_j\hat{a}_{j+\Delta l}\rangle}e^{\left(iz\Delta l^2/2-izN'_L\Delta l\right)}\frac{1-e^{i\left(2N'_L-\Delta l +1\right)z\Delta l}}{1-e^{iz\Delta l}} \label{eq:correction}
\end{align}
The total site number $N_L$ and $\overline{\langle\hat{a}^\dagger_j\hat{a}_{l}\rangle}$ are obtained 
by fitting the long-range atomic correlation $\sum\langle\hat{a}^\dagger_j\hat{a}_{j+\Delta l}\rangle$ 
by use of the experimental data with various lattice separation $\Delta l =1,2,3,4$ and several TOFs.
In our experiment to measure the ensemble average of the kinetic term, the TOFs were 14 and 18 ms.
Figure~\ref{fig:correction} shows our typical measured long-range atomic correlation $\sum\langle\hat{a}^\dagger_j\hat{a}_{j+\Delta l}\rangle$ and the fitting curves obtained using Eq. (\ref{eq:correction}).
Finally, by fitting $\overline{\langle\hat{a}^\dagger_j\hat{a}_{l}\rangle}$ with Eq.~(\ref{eq:coherent}), we obtain the coherence length $\xi$.

\begin{figure}[tb]
	\includegraphics[width=9cm]{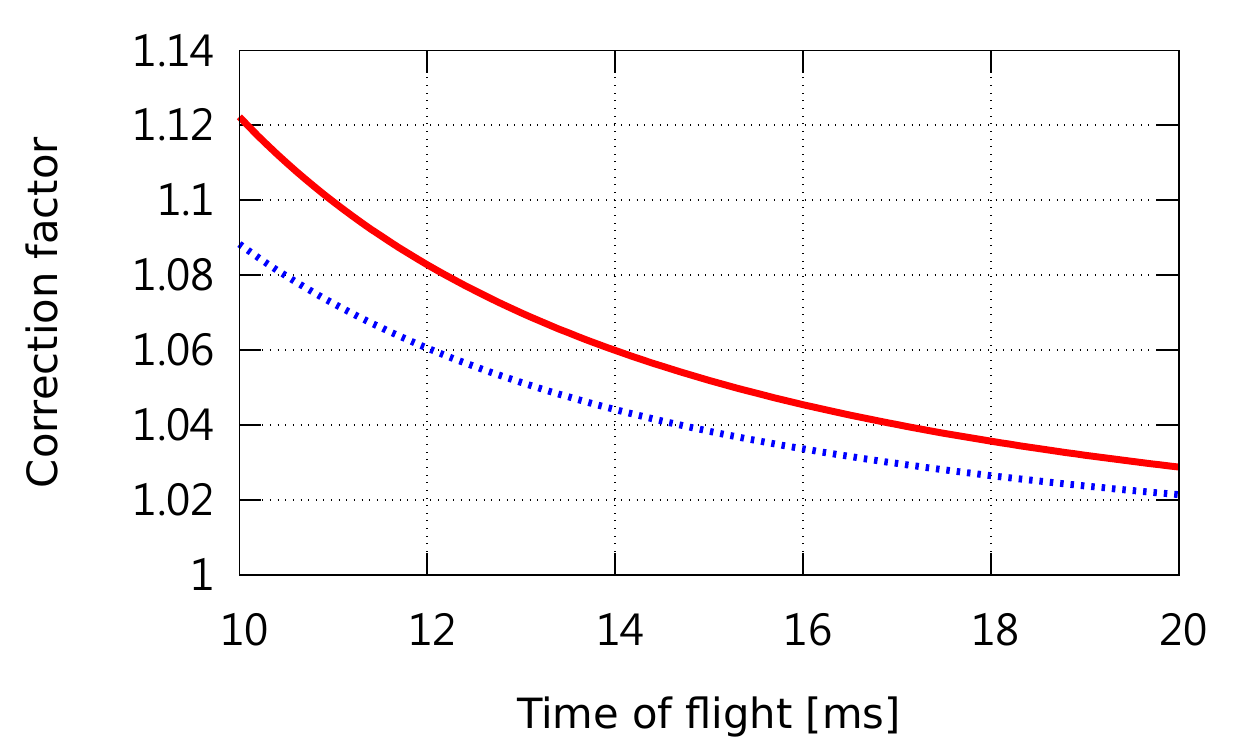}
	\caption{(color online) Correction factors as a function of time of flight.
The red solid line shows the correction factor $C_1$ calculated using the method we used for data analysis.
Here we used $N_L=81$.
For reference, we show another correction factor $C'$ based on the assumption of $\langle \hat{a}^{\dagger}_j \hat{a}_l\rangle=\sqrt{n_jn_l}\exp\left(-(x_j-x_l)/\xi'\right)$ is shown as the blue dotted line. We assumed $\xi'=10d_{\text{lat}}$, which is estimated by use of fitting results using our experimental data.
\label{fig:correction2}}
\end{figure}

The correction factor $C_1$ is shown in Fig.~\ref{fig:correction2} as the solid red line using our experimental parameters.
The deviation by the Fresnel effect is estimated to be about 6\% (4\%) for 14 ms (18 ms) TOF.
It is to be noted that the correction factor also depends on total atom size $N_L$ and monotonically decrease with the limit $N_L\to0$.

Here we assume that the atom correlation have the average value of $\overline{\langle\hat{a}^\dagger_j\hat{a}_{l}\rangle}$ and unique atom density, which is valid for a Mott insulating case.
However, this is not a unique possible assumption and the calculated correction factor depends on models.
For reference, we calculated a correction factor $C'$ based on the assumption of $\langle \hat{a}^{\dagger}_j \hat{a}_l\rangle=\sqrt{n_jn_l}\exp\left(-(x_j-x_l)/\xi'\right)$ and is shown in Fig.~\ref{fig:correction2} as the blue dotted line, where we assumed that density distribution the Gaussian function and the $\xi'$ is a correlation length~\cite{Braun2015}.
In this model, because of large atom density around the center of the trap, effective size of atoms are small compared to the model with unique atom density, and therefore small correction factor obtained.
Apart from non-realistic cases (namely, atom density around the edge of the trap is large compared to the one in the center), atoms with unique density have the largest effective size, and it results in the largest correction factor. 
These estimations show that the maximum of the correction factor may obtain from the model with unique atom density.
Therefore we also use estimated difference between the values before and after the correction as systematic errors in order to cover the uncertainty of atoms density distribution in the trap.
In a coexistence case of SF-Mott phase, the correction are expected within the systematic errors.

\section{Measurement of visibility, width, and coherence length} \label{ap:other}
\begin{figure*}
	\includegraphics[width=17cm]{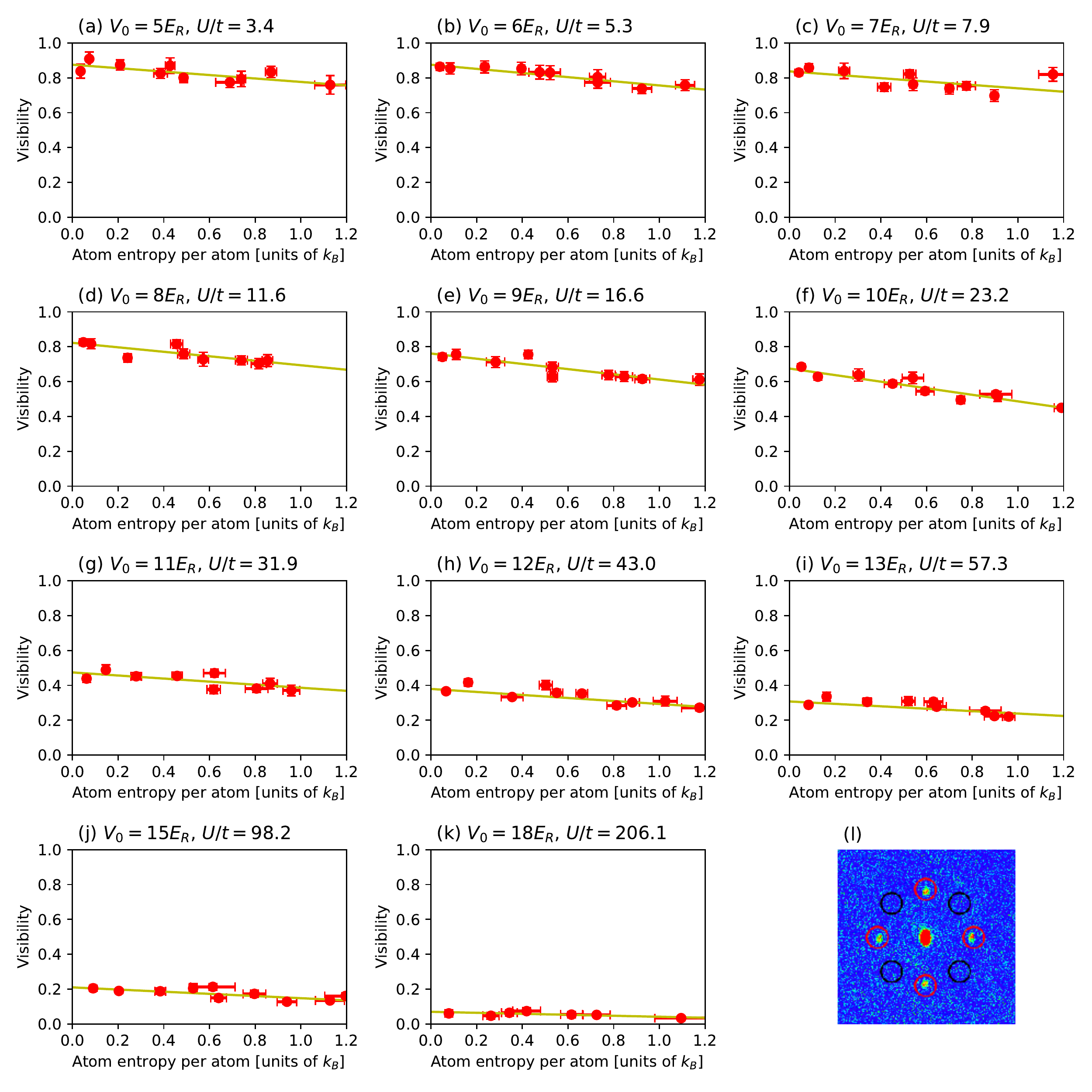}
	\caption{(color online) (a--k) Visibilities as functions of atomic entropy. 
		The error bars show standard errors. 
		The yellow lines are a guide for the eye. 
		(l) Interference pattern. The maxima of the interference pattern peak appear at the first peaks (red circles), and the minima peaks appear at diagonal with the same distance from the central peak (black circles).
		The sum of the signals in the red (black) circles is $n_{\max}$ ($n_{\min}$) (see text for details).\label{fig:visibility}}
\end{figure*}

\begin{figure*}
	\includegraphics[width=17cm]{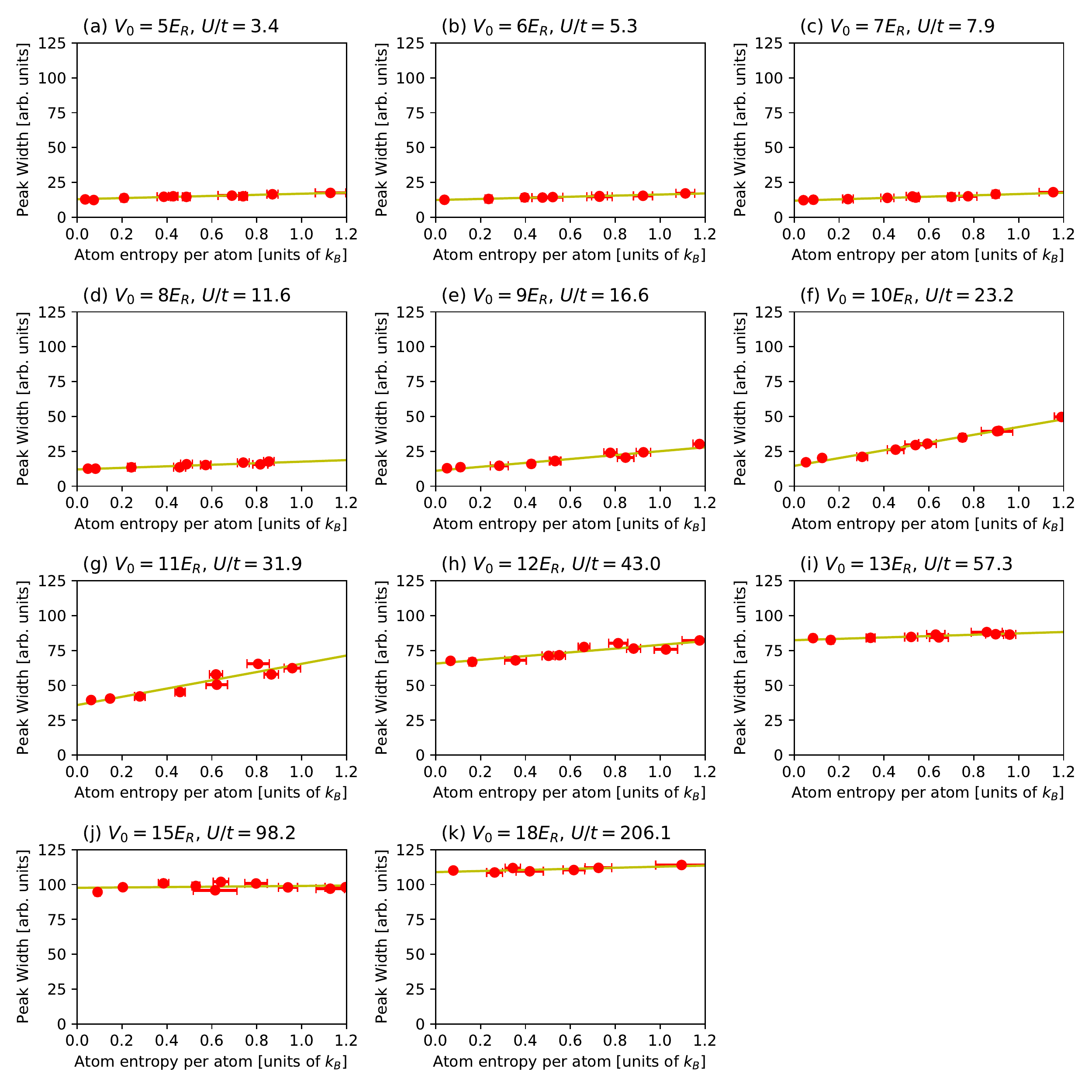}
	\caption{(color online) Widths of central peaks as functions of atomic entropy. The yellow lines are a guide for the eye. The error bars show standard errors.\label{fig:width}}
\end{figure*}

The widely used experimental observables from the TOF images are the visibility (Fig.~\ref{fig:visibility}) and peak width (Fig.~\ref{fig:width}).
Figures~\ref{fig:visibility}(a--k) show the visibilities as functions of atomic entropy.
The visibility $\mathcal{V}$ is defined as~\cite{Gerbier2005, Gerbier2007}
\begin{equation}
\mathcal{V}=\frac{n_{\max}-n_{\min}}{n_{\max}+n_{\min}},
\end{equation}
where $n_{\max}$ is the maximum density at the first interference peak. The minimum density $n_{\min}$ is measured at the same distance, but in a diagonal direction from the central peak (see also, Fig.~\ref{fig:visibility}(l)).
It is clearly apparent that the visibility is large at small lattice depth and decreases as the lattice depth increases.
The dependence of the visibility on the atomic entropy is small.

Figure~\ref{fig:width} shows the widths of the central peaks as functions of the atomic entropy.
The central peak width is one of the most commonly used parameters to evaluate the phase coherence.
If the TOF is sufficiently long to neglect the Fresnel effect (see Eq. (\ref{eq:TOF1})),
the structure factor $S(\mathbf{k}=0)$ is
\begin{eqnarray}
S(\mathbf{k}=0)&=&\sum_{j,l}\langle \hat{a}^{\dagger}_j \hat{a}_l \rangle \nonumber \\
&=&\langle\hat{a}^{\dagger}_0\hat{a}_0\rangle+\langle\hat{a}^{\dagger}_0\hat{a}_1\rangle+\langle\hat{a}^{\dagger}_0\hat{a}_2\rangle +\cdots \nonumber \\
& &+\langle\hat{a}^{\dagger}_1\hat{a}_0\rangle+\langle\hat{a}^{\dagger}_1\hat{a}_1\rangle+ \langle\hat{a}^{\dagger}_1\hat{a}_2\rangle +\cdots \nonumber\\
& & +\cdots \nonumber\\
&=&\langle(\hat{a}^\dagger_0+\hat{a}^\dagger_1+\cdots)(\hat{a}_0+\hat{a}_1+\cdots)\rangle \nonumber\\
&=&\langle(\hat{c}^\dagger(\textbf{k}=0)\hat{c}(\textbf{k}=0)\rangle \nonumber\\
&=&N_L |\phi|^2,
\end{eqnarray}
where $\phi$ is the wavefunction of the SF component.
It is naturally expected that a larger phase coherence corresponds to a sharper peak width.
One can clearly see that the central peak is sharp at shallow lattice depth and increases with lattice depth.
The dependence of the peak width on the atomic entropy is small.

While these measurements have been standard methods in the study of the SF-MI transition,
the new internal energy measurements of the Bose-Hubbard system demonstrated in this work provide a useful method of investigating the SF-MI transition, as shown in the main text.

\begin{figure*}
	\includegraphics[width=17cm]{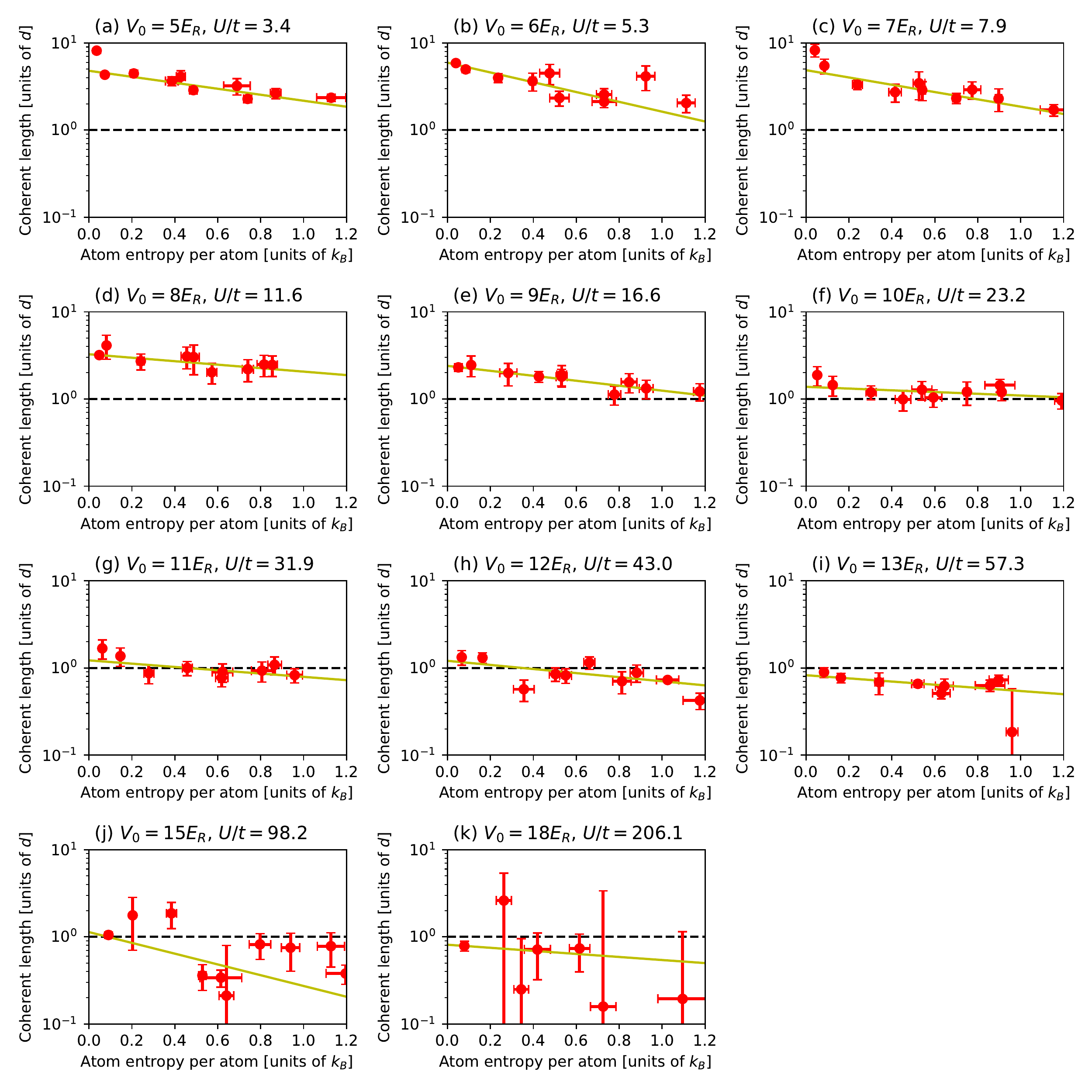}
	\caption{(color online) Coherence lengths $\xi$ as functions of atomic entropy. 
		The Y-axis is the log scale and $d$ is the lattice spacing. 
		The yellow lines are guides for the eye. The error bars show standard errors.\label{fig:coherence}}
\end{figure*}

Figure~\ref{fig:coherence} shows the coherence lengths $\xi$ as functions of the atomic entropy.
Our Fourier transformation method enables us to consider the long-range atomic correlation of more than just the nearest-neighbor sites.
Here, $\xi$ is defined as~\cite{Braun2015}
\begin{equation}
\langle \hat{a}^{\dagger}_j \hat{a}_l\rangle=\sqrt{n_j}\sqrt{n_l}\exp\left(-\frac{|\mathbf{r}_j-\mathbf{r}_l|}{\xi}\right),\label{eq:coherent}
\end{equation}
where $n_j$ is the atomic density at site $j$.
The value of $\xi$ is obtained by fitting Eq.~(\ref{eq:coherent}) to our measured ensemble average of the long-range atomic correlation $\overline{\langle\hat{a}^\dagger_j\hat{a}_{l}\rangle}$ (see Appendix~\ref{ap:TOF}).
Note that $\xi$ is large at a small lattice depth and decreases with increased lattice depth.
As expected, $\xi$ is near one lattice spacing around the quantum critical point ($s_c=10.6$ for $n=1$).
This behavior also shows the quantum phase transition between SF and MI.

\section{Atom-number-projection spectroscopy procedure} \label{ap:interaction}
We used the transition from the ($6s^2$) \SSS state to the ($6s6p$)\PPP ($m_J=0$) state for high-resolution spectroscopy. 
Neither the \SSS nor the \PPP ($m_J=0$) state is sensitive to magnetic fields, because \Yb{174} lacks a nuclear spin; this enabled us to obtain narrow spectra in the absence of inhomogeneous broadening resulting from an external magnetic field.

Light for the excitation was generated through frequency doubling of an external-cavity laser diode at 1014 nm, locked to an ultralow expansion cavity, which had slow-frequency drift with a typical rate of approximately 1 kHz/h.
The linewidth of the excitation laser was less than 1 kHz.

After atom projection to a large optical depth of $15$$E_R$, as described above, we applied an excitation pulse. 
The pulse width was 0.3 ms.
The incident power was approximately 100 $\mu$W and the beam waist was approximately 50 $\mu$m.
The intensity was $2.5$ W/$\text{cm}^2$ and the Rabi frequency was approximately 0.3 kHz.
To excite the \SSS-\PPP ($m_J=0$) transition, the excitation light propagating along the Y-axis was polarized along the Z-axis and we applied a magnetic field of $100$ mG in the -X+Y direction.

After applying the excitation light, we removed the remaining atoms in the \SSS state with a light that resonated with the \SSS-($6s6p$)\PPPS transition for 0.3 ms.
Then, atoms in the \PPP state were transferred to the \SSS state with two repumping lights resonant with the ${}^{3}\text{P}_{2}$ - ($6s7s$)${}^{3}\text{S}_{1}$ and ($6s6p$)${}^{3}\text{P}_{0}$-($6s7s$)${}^{3}\text{S}_{1}$ transitions.
Finally, the number of atoms in the \SSS state was measured using a fluorescence imaging technique employing a magneto-optical trap with the \SSS-\PPPS transition.

Typical spectra have already been shown in Figs~\ref{fig:projection} (b) and \ref{fig:spectra}.
Our spectra were obtained after projection into the lattice depth of $15$$E_R$; thus, the positions of each peak relative to the $n=1$ peak were fixed.
Therefore, our spectra covered four peaks corresponding to $n=1 \cdots 4$.

The spectral areas were obtained by fitting using a sinc function, because our excitation light pulse was rectangular and the resulting broadening from Fourier transformation of the rectangular function was dominant.
The correction factors from the reduced Rabi frequencies and finite lifetimes of atoms in the \PPP state ($n$ = $2,3,4$) were considered.
We took three or more spectra and calculated the atomic distribution for each one; then, these data were averaged.
To save time on our experiment, only several data points in the vicinities of peaks were taken.
A period of approximately 20 min was required to obtain one spectrum and the long-term drift was negligible for the time scale.

\section{Possible Temperature Estimation from Energy Measurements} \label{ap:temperature}
Although the atomic temperature in an optical trap without an optical lattice can be easily measured using a TOF method, 
the atomic temperature in an optical lattice is estimated only indirectly through comparison of the experimental results and theoretical calculation.
In the higher-temperature region, estimation of the atom temperature from the {\it in situ} atom distribution~\cite{DeMarco2009} and spin-gradient thermometry~\cite{Weld2009} has been demonstrated, as well as use of quantum gas microscopy~\cite{DeMarco2009}.

Alternatively, however, if the total internal energies are measured experimentally, one can determine the temperature $T$ using the thermodynamic relation
\begin{equation}
	T=\frac{\partial E}{\partial S}, \label{eq:thermo}
\end{equation}
where $E$ is the total internal energy and $S$ is the atomic entropy.
%
\begin{figure}[tb]
	\includegraphics[width=9cm]{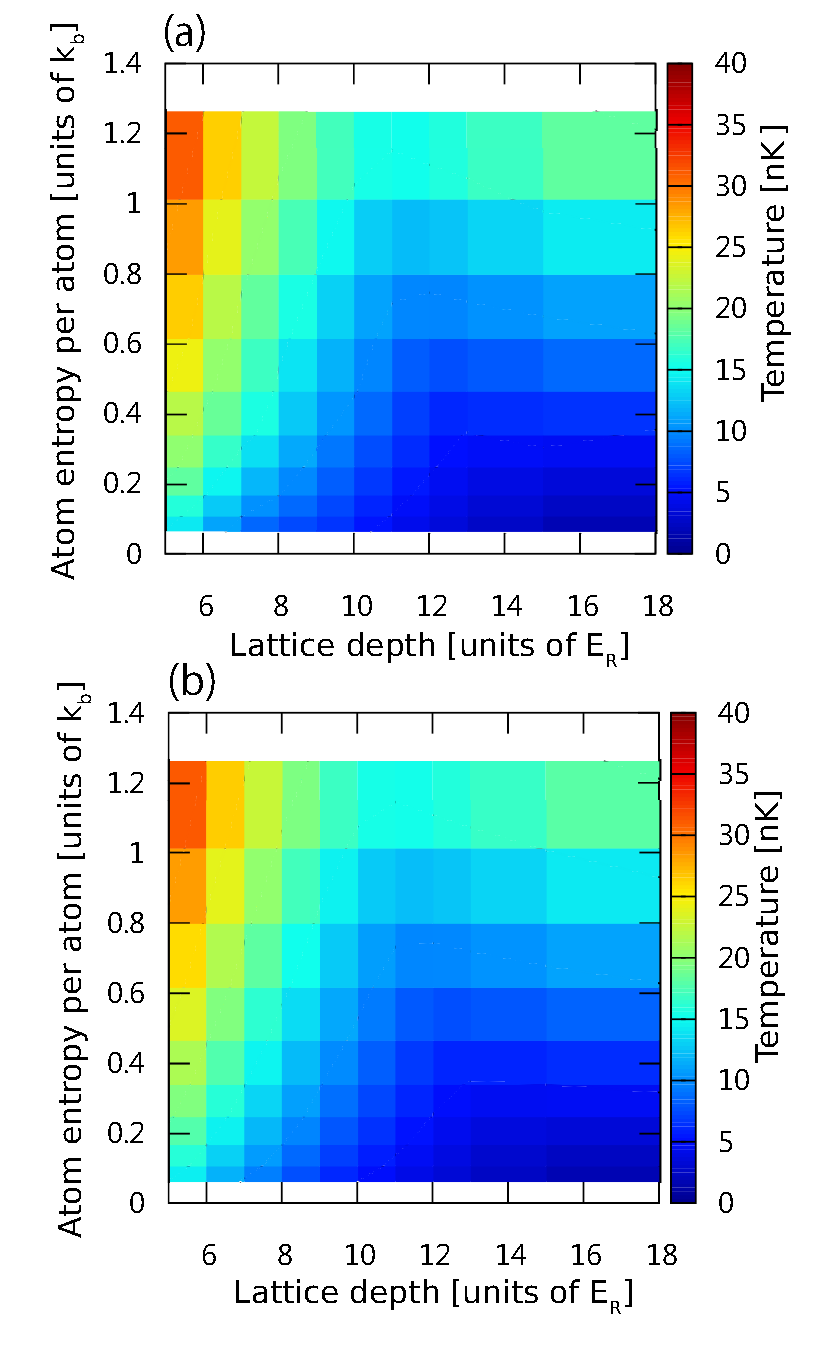}
	\caption{(color online) (a) Estimated temperature using relation $T=\partial E/\partial S$. 
The ensemble averages of the kinetic, interaction, and potential terms for estimation were obtained from the Gutzwiller approximation. (b) Temperature value obtained using numerical calculations with Gutzwiller approximation. 
For both cases, the total atom number was $1.4\times10^4$ and we used the same trap conditions as in the main paper.\label{fig:estimation}}
\end{figure}
%
In our experiment, evaluation of the potential energy is difficult, as mentioned.

Figure~\ref{fig:estimation} (a) shows the temperature estimated using the relation $T=\partial E/\partial S$.
The ensemble averages of the kinetic, interaction, and potential terms for estimation were obtained from the Gutzwiller approximation. 
This result is consistent with the temperature directly obtained using numerical calculation with the Gutzwiller approximation and shown in  Fig.~\ref{fig:estimation} (b). 

The contribution of the potential term comes from the trap potentials and the Gaussian envelope of the optical lattice lasers.
This is because both external potentials are quadratic terms with respect to the lattice index; that is, $V_j=m\omega^2(V_0)(\textbf{r}_j-\textbf{r}_0)^2/2$, where $\omega(V_0)$ is the overall (mean) trap frequency as a function of lattice depth $V_0$ and $\textbf{r}_0$ is the central position of the overall external potential.

Therefore, the Bose-Hubbard-model Hamiltonian is expressed as
\begin{align}
\hat{\mathcal{H}}=&-t(V_0)\sum_{\langle j, l \rangle}\left(\hat{a}^{\dagger}_j \hat{a}_l +h.c.\right)+\frac{U(V_0)}{2}\sum_{j} \hat{a}^{\dagger}_j\hat{a}^{\dagger}_j\hat{a}_j\hat{a}_j \nonumber \\
&+P(V_0)\sum_{j}(\textbf{r}_j-\textbf{r}_0)^2\hat{a}^{\dagger}_j\hat{a}_j-\mu N, \label{eq:Hamiltonian2}
\end{align}
where $N$ is the total atom number and $P(V_0)=m\omega^2(V_0)/2$.
Note that $t(V_0)$, $U(V_0)$, and $P(V_0)$ are known functions that depend only on $V_0$.

We apply the Hellmann--Feynman theorem~\cite{Hellmann,Feynman1939} to the ensemble average of the Hamiltonian $E(V_0, S_{\text{OL}})=\langle\hat{H}\rangle$:
\begin{equation}
\frac{\partial}{\partial V_0}E(V_0, S_{\text{OL}})=\left\langle\frac{d\hat{H}}{dV_0}\right\rangle, \label{eq:H-F}
\end{equation}
where $S_{\text{OL}}$ is the atomic entropy in the optical lattice and 
\begin{align}
\frac{\partial}{\partial V_0}E(V_0, S_{\text{OL}})=& -\frac{dt(V_0)}{dV_0}
\sum_{\langle j, l \rangle}\langle\hat{a}^{\dagger}_j \hat{a}_l\rangle \nonumber \\
&+\frac{1}{2}\frac{dU(V_0)}{dV_0}\sum_{j}\langle\hat{a}^{\dagger}_j\hat{a}^{\dagger}_j\hat{a}_j\hat{a}_j\rangle \nonumber \\
&+\frac{dP(V_0)}{dV_0}\sum_{j}(\textbf{r}_j-\textbf{r}_0)^2\langle\hat{a}^{\dagger}_j\hat{a}_j\rangle \nonumber \\
=&-\frac{dt(V_0)}{dV_0}
K(V_0,S_{\text{OL}}) \nonumber \\
&+\frac{1}{2}\frac{dU(V_0)}{dV_0}G(V_0,S_{\text{OL}}) \nonumber \\
&+\frac{dP(V_0)}{dV_0}L(V_0,S_{\text{OL}}).
\end{align}
The values of $K(V_0,S_{\text{OL}})$, $G(V_0, S_{\text{OL}})$, and $L(V_0, S_{\text{OL}})$ are experimentally observed and given by
\begin{align}
K(V_0,S_{\text{OL}}) &= \sum_{\langle j, l \rangle}\langle\hat{a}^{\dagger}_j \hat{a}_l\rangle \\
G(V_0, S_{\text{OL}})&=\sum_{j}\langle\hat{a}^{\dagger}_j\hat{a}^{\dagger}_j\hat{a}_j\hat{a}_j\rangle \\
L(V_0, S_{\text{OL}})&=\sum_{j}(\textbf{r}_j-\textbf{r}_0)^2\langle\hat{a}^{\dagger}_j\hat{a}_j\rangle,
\end{align}
and we omit $\langle \hat{\cdot} \rangle$ for simplicity in this section (that is, $K$ instead of $\langle\hat{K}\rangle$).
Therefore,
\begin{align}
\frac{\partial T(V_0, S_{\text{OL}})}{\partial V_0}=&\frac{\partial}{\partial V_0}\frac{\partial}{\partial S_{\text{OL}}}E(V_0, S_{\text{OL}}) \nonumber \\
=&\frac{\partial}{\partial S_{\text{OL}}}\frac{\partial}{\partial V_0}E(V_0, S_{\text{OL}}) \nonumber \\
=& -\frac{dt(V_0)}{dV_0}
\frac{\partial K(V_0, S_{\text{OL}})}{\partial S_{\text{OL}}} \nonumber \\
&+\frac{1}{2}\frac{dU(V_0)}{dV_0}\frac{\partial G(V_0, S_{\text{OL}})}{\partial S_{\text{OL}}} \nonumber \\
&+\frac{dP(V_0)}{dV_0}\frac{\partial L(V_0, S_{\text{OL}})}{\partial S_{\text{OL}}},
\end{align}
where $T(V_0, S_{\text{OL}})$ is the atomic temperature.

Even if the ensemble average of the potential terms is unavailable, the atomic temperature can be estimated. 
To demonstrate this, we consider the normalized operator $\hat{H}'=\hat{H}/P(V_0)$ and its ensemble average.
\begin{align}
\frac{\partial }{\partial V_0}\left(\frac{E(V_0,S_{\text{OL}})}{P(V_0)}\right)=&\left\langle\frac{d}{dV_0}\left(\frac{\hat{H}}{P(V_0)}\right)\right\rangle \nonumber \\
=& -\frac{d}{dV_0}\left(\frac{t(V_0)}{P(V_0)}\right) \sum_{\langle j, l \rangle}\langle\hat{a}^{\dagger}_j \hat{a}_l\rangle \nonumber \\
&+\frac{1}{2}\frac{d}{dV_0}\left(\frac{U(V_0)}{P(V_0)}\right)\sum_{j}\langle\hat{a}^{\dagger}_j\hat{a}^{\dagger}_j\hat{a}_j\hat{a}_j\rangle \nonumber \\
=&-K(V_0,S_{\text{OL}})\frac{d}{dV_0}\left(\frac{t(V_0)}{P(V_0)}\right) \nonumber \\
&+\frac{1}{2}G(V_0,S_{\text{OL}})\frac{d}{dV_0}\left(\frac{U(V_0)}{P(V_0)}\right). \label{eq:Tdepend1}
\end{align}
In contrast,
\begin{equation}
\left\langle\hat{H}'\right\rangle=\left\langle\frac{\hat{H}}{P(V_0)}\right\rangle=\frac{\left\langle\hat{H}\right\rangle}{P(V_0)}=\frac{E(V_0,S_{\text{OL}})}{P(V_0)}.
\end{equation}
Because
\begin{equation}
\frac{T(V_0, S_{\text{OL}})}{P(V_0)}=\frac{\partial}{\partial S_{\text{OL}}}\frac{E(V_0,S_{\text{OL}})}{P(V_0)},
\end{equation}
the dependence of $T$ on $V_0$ is
\begin{align}
\frac{\partial}{\partial V_0}\left(\frac{T(V_0, S_{\text{OL}})}{P(V_0)}\right)=&\frac{\partial}{\partial V_0}\frac{\partial }{\partial S_{\text{OL}}}\left(\frac{E(V_0,S_{\text{OL}})}{P(V_0)}\right) \nonumber \\
=&\frac{\partial }{\partial S_{\text{OL}}}\frac{\partial}{\partial V_0}\left(\frac{E(V_0,S_{\text{OL}})}{P(V_0)}\right).
\end{align}
Using Eq.~(\ref{eq:Tdepend1}),
\begin{align}
\frac{\partial}{\partial V_0}\left(\frac{T(V_0, S_{\text{OL}})}{P(V_0)}\right)=&-\frac{\partial K(V_0,S_{\text{OL}})}{\partial S_{\text{OL}}}\frac{d}{dV_0}\left(\frac{t(V_0)}{P(V_0)}\right) \nonumber \\
&+\frac{1}{2}\frac{\partial G(V_0,S_{\text{OL}})}{\partial S_{\text{OL}}}\frac{d}{dV_0}\left(\frac{U(V_0)}{P(V_0)}\right). \label{eq:Tdepend2}
\end{align}
Equation~(\ref{eq:Tdepend2}) shows that we must obtain the dependencies of $K(V_0,S_{\text{OL}})$ and $G(V_0,S_{\text{OL}})$ on the atom entropy $S_{\text{OL}}$ because $t(V_0)$, $U(V_0)$, and $P(V_0)$ are all known functions.
Therefore, direct measurement of the potential term $L(V_0,S_{\text{OL}})$ is not necessary to estimate the atomic-temperature dependence.
When we know the absolute atomic temperature $T(V'_0,S_{\text{OL}})$ at a certain lattice depth $V'_0$, 
we can estimate the other absolute atomic temperatures through integration of Eq.~(\ref{eq:Tdepend2}).



%

\end{document}